\documentclass{aa}
\usepackage[varg]{txfonts}
\usepackage[font=footnotesize,labelfont={bf,sc,small},figurename=Fig.]{caption}
\usepackage[colorlinks,allcolors=blue]{hyperref}
\usepackage{subfig}
\usepackage{rotate}
\usepackage{appendix}

\usepackage{varioref}
\labelformat{figure}{Fig.~#1}
\labelformat{section}{Sect.~#1}
\labelformat{subsection}{Sect.~#1}
\labelformat{subsubsection}{Sect.~#1}
\labelformat{appendix}{Appendix~#1}
\labelformat{table}{Table~#1}
\usepackage{natbib}
\bibpunct{(}{)}{;}{a}{}{,} 

\usepackage{multirow}
\begin{document}

\title{An X-ray activity cycle on the young solar-like star $\epsilon \ \rm Eri$dani}

\author{M. Coffaro\inst{\ref{inst1}}
\and B. Stelzer\inst{\ref{inst1},\ref{inst2}}
\and S. Orlando\inst{\ref{inst2}}
\and J. Hall \inst{\ref{inst3}}
\and T.S. Metcalfe\inst{\ref{inst4}}
\and U. Wolter\inst{\ref{inst5}}
\and M. Mittag\inst{\ref{inst5}}
\and J. Sanz-Forcada\inst{\ref{inst6}}
\and P.C. Schneider\inst{\ref{inst5}}
\and L. Ducci\inst{\ref{inst1}}
} 

\offprints{M. Coffaro}

\institute{Institut f\"ur Astronomie und Astrophysik T\"ubingen (IAAT), Eberhard-Karls Universit\"at T\"ubingen, Sand 1, D-72076, Germany \email{coffaro@astro.uni-tuebingen.de}\label{inst1}
\and INAF - Osservatorio Astronomico di Palermo, Piazza del Parlamento 1, I-90134 Palermo, Italy\label{inst2}
\and Lowell Observatory, 1400 West Mars Hill Road, Flagstaff, AZ 86001\label{inst3}
\and Space Science Institute, 4765 Walnut Street, Suite B, Boulder, CO 80301, USA \label{inst4}
\and Hamburger Sternwarte, Universit\"at Hamburg, Gojenbergsweg 112, 21029, Hamburg, Germany \label{inst5}
\and Departamento de Astrof\'isica, Centro de Astrobiolog\'ia (CSIC-INTA), ESAC Campus, Camino bajo del Castillo s/n, E-28692 Villanueva de la Ca\~nada, Madrid, Spain\label{inst6}. 
}

\date{Received 8 August 2019 / Accepted 20 February 2020}

\abstract {Chromospheric Ca\,II activity cycles are frequently found in late-type stars,
but there have been no systematic programs to search for their coronal X-ray counterparts. The typical time
scale of Ca\,II activity cycles goes from years to decades. Therefore, long-lasting missions are needed to detect
the coronal counterparts. \textit{XMM-Newton} has so far detected X-ray cycles in five
stars. A particularly intriguing question is at
what age (and at what activity level) X-ray cycles set in. To this end, in 2015 we started the X-ray monitoring of the young solar-like star $\epsilon \ \rm Eridani$, observed previously twice in 2003 and in early 2015 by \textit{XMM-Newton}.
With an age of $440$ Myr, it is one of the youngest solar-like stars with a known chromospheric Ca\,II cycle. We collected the most recent Mount Wilson S-index
data available for $\epsilon \ \rm Eridani$, starting from 2002, including previously unpublished data. We found that the Ca\,II cycle lasts $2.92 \pm 0.02$ yr, in
agreement with past results. From the long-term \textit{XMM-Newton} lightcurve, we find clear and systematic X-ray 
variability of our target, consistent with the chromospheric Ca\,II cycle. The average X-ray luminosity results to be $2 \times 10^{28} \rm erg/s$, with an amplitude that is only a factor $2$ throughout the cycle.  We apply a new method to describe the evolution of the coronal emission measure distribution of $\epsilon \ \rm Eridani$ in terms of solar magnetic structures: active regions, cores of active regions and flares covering the stellar surface at varying filling fractions. Combinations of these three types of magnetic structures can describe the observed X-ray emission measure of $\epsilon \ \rm Eridani$  only if the solar flare emission measure distribution is restricted to events in the decay phase. The interpretation is that flares in the corona of $\epsilon \ \rm Eridani$ last longer than their solar counterparts. We ascribe this to the lower metallicity of $\epsilon \ \rm Eridani$. Our analysis revealed also that the X-ray cycle of $\epsilon \ \rm Eridani$ is strongly dominated by cores of active regions. The coverage fraction of cores throughout the cycle changes by the same factor as the X-ray luminosity. The maxima of the cycle are characterized by a high percentage of covering fraction of the flares, consistent with the fact that flaring events are seen in the corresponding short-term X-ray lightcurves predominately at the cycle maxima. The high X-ray emission throughout the cycle of $\epsilon \ \rm Eridani$ is thus explained by the high percentage of magnetic structures on its surface.}

\maketitle


\section{Introduction}
\label{sec:intro}
Stellar dynamos maintain the magnetic field of late-type stars on long-timescales through a periodic field reversal. The dynamo cycle comprises a phase in which the interior field rises to the surface where it forms magnetic loop structures. As yet poorly determined, magnetic heating processes produce a positive temperature gradient above the photosphere, and ensuing characteristic emission from plasma at $T \sim \ 10000$ K to few $10^6$ K. As the surface coverage with these high-activity regions changes throughout the dynamo cycle, an \textit{activity cycle} of the same length is associated with the magnetic field cycle. Observations of such activity cycles can, thus, be used as proxies for the stellar dynamo.

In the chromosphere, the monitoring of H and K emission lines of Ca\,II is the most widely used indicator for activity cycles. In a dedicated monitoring program at the Mt. Wilson telescope \citep{1968ApJ...153..221W, 1978ApJ...226..379W}, the ubiquitous existence of cyclic stellar activity in cool stars was revealed: about $60 \%$ of the main-sequence stars with spectral types F and M present variability of the Ca\,II H\&K Mount Wilson S-index, $S_{\rm MWO}$, \citep{1995ApJ...438..269B}, showing periodicities from $2$ yr up to $20$ yr. Finding the X-ray counterpart of activity cycles, i.e. their manifestation in the stellar corona, is still challenging. Due to the typical cycle length of years to decades, long-term X-ray monitoring campaigns of activity cycles are unfeasible for a significant sample. Such studies require a long-lived X-ray mission, as problems of cross-calibration can arise when different telescopes are used, because the data would have different wavelength coverage and responses. 

Up to date, \textit{XMM-Newton} has detected X-ray activity cycles in five stars\footnote{{Among the stars observed multiple times with XMM-Newton, other two stars show variability in the X-ray waveband, compatible with the chromospheric cycles observed during the Mount Wilson project. These stars are $\tau$ Bo\"otis \citep{2017A&A...600A.119M}, with a cycle period of $\sim 4$ months, the shortest period observed, and the third companion of the stellar system $\alpha$ Cen, Proxima Cen, with an evidence of an activity cycle recently found \citep{10.1093/mnras/stw2570}.}}. Four of these stars are part of stellar systems: 61 Cyg A \citep{2006A&A...460..261H,2012A&A...543A..84R}, $\alpha$ Cen A and $\alpha$ Cen B \citep{2012A&A...543A..84R, 10.1093/mnras/stw2570} and HD 81809 \citep{2008A&A...490.1121F, 2017A&A...605A..19O}. They are old stars with ages of several Gyr and they show long X-ray cycle periods, from $8 \ \rm yr$ to $\sim 19 \ \rm yr$. \citet{2008A&A...490.1121F} and \citet{2017A&A...605A..19O} have hypothesized that the stellar X-ray activity of HD 81809 comes from the primary component of the binary system. This statement was questioned in the literature \citep{2018ApJ...855...75R}, but the main obstacle in constraining such systems comes from the geometrical configuration of the system that is not spatially resolved \citep{2018ApJ...866...80E}. 

The fourth star monitored by \textit{XMM-Newton}, $\iota$ Horologii, has different characteristics from the others: it is relatively young with an age of $\sim 600 \ \rm Myr$ and a cycle period of $1.6 \ \rm yr$, the shortest X-ray cycle measured until now \citep{2013A&A...553L...6S}. The detection of an X-ray cycle in $\iota$ Hor has shown that coronal cycles can be also found in young stars. This triggers the question at which age and at which activity level X-ray cycles set in. To this end, we need well-selected targets such as $\iota \ \rm Hor$, i.e. young and active solar-like stars with short chromospheric cycle period, enabling an X-ray detection in reasonable time-span. 

$\epsilon$ Eridani ($\epsilon \ \rm Eri$; HD 22049) is a young {roughly} solar-type dwarf star with spectral type K2V. It is at a distance of $3.2 \ \rm pc$ \citep{2007A&A...474..653V}, has an age of $440 \ \rm Myr$ \footnote{In the literature the age of $\epsilon \ \rm Eri$ is estimated to be between $200$~Myr and  $930 \ \rm Myr$ \citep{2004AN....325....3F, 2000ApJ...533L..41S}. We adopt here the age of $440 \ \rm Myr$ found by \citet{2007ApJ...669.1167B} through gyro-chronology.} and a radius of $0.74 \ R_{\odot}$ \citep{2007ApJ...669.1167B, 2004A&A...426..601D}. It hosts two Jupiter-like planets with a semimajor axis of $3.4$ AU and $40 $ AU respectively \citep{2000ApJ...544L.145H, 2002ApJ...578L.149Q}. $\epsilon \ \rm Eri$ was part of the Mt. Wilson project, and the monitoring of its chromospheric emission started in 1968. These data were first published by \citet{1995ApJ...441..436G}, finding a cycle period of approximately $5$ yr. Later \citet{2013ApJ...763L..26M} combined the Mount Wilson data with more recent observations carried out at other observatories.
They found two periodicities in the long-term lightcurve of the S-index: $2.95 \pm 0.03$ yr and $12.7 \pm 0.3$ yr. 

The clear evidence for a short chromospheric cycle together with its youth, have led us to start an \textit{XMM-Newton} X-ray monitoring program of $\epsilon \ \rm Eri$. Here we report the first detection of a $\sim \ 3$ yr X-ray cycle. 

In \ref{sec:obs} we present the most recent Ca\,II $S_{\rm MWO}$-index data and our X-ray monitoring campaign of $\epsilon \ \rm Eri$, together with the description of the data reduction.
In \ref{sec:sim} we present a novel method in which we describe the X-ray emission and the evolution of the coronal cycle of $\epsilon \ \rm Eri$ in terms of varying solar-like magnetic structures. It consists in comparing the emission measure distributions (EMDs) of the magnetic structures observed on the Sun to that of $\epsilon \ \rm Eri$. \citet{2008A&A...490.1121F} and \citet{2017A&A...605A..19O} laid the foundations for this technique in application to HD~81809. Here, the high-quality spectra of $\epsilon \ \rm Eri$ allow us here a much more detailed study in which we also refine the method. 
In \ref{sec:disc} and \ref{sec:conc} we discuss our results and we give our conclusions.

\section{Observations and data analysis}
\label{sec:obs}
\subsection{Ca\,II H\&K data}
\label{sub:calcio}
As support of our search for an X-ray activity cycle, we collected Ca\,II {$S_{\rm MWO}$-index measurements} of $\epsilon \ \rm Eri$.  $\epsilon \ \rm Eri$ was observed within the Mount Wilson project, from the late 60s to early 90s. After the Mount Wilson project the monitoring of the Ca\,II H\&K lines continued at other observatories. The data we take into account in this article for $\epsilon \ \rm Eri$ were obtained from 2002 {to late 2018} such as to cover the full time-span of the existing \textit{XMM-Newton} observations of $\epsilon \ \rm Eri$. 

\begin{figure}[!htbp]
\centering
\includegraphics[width=\hsize]{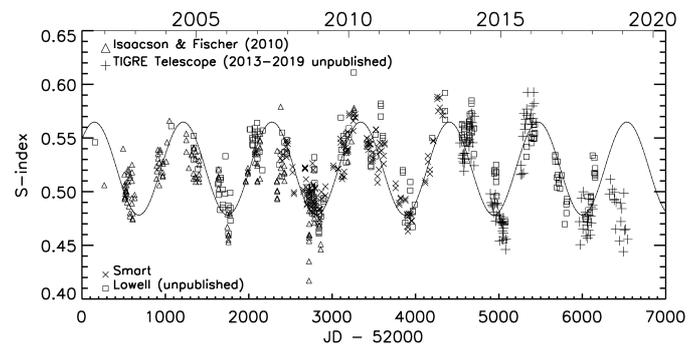}
\caption{Ca\,II Mount Wilson S-index of $\epsilon \ \rm Eri$. The data set covers a period from 2002 to late 2018 and it comes from four different instruments: \textit{SMARTS} 1.5 m telescope \textit{(cross-symbols)}, \textit{SSS} at Lowell Observatory \textit{(squared-symbols)}, \textit{HIRES} at the Keck Observatory \textit{(triangle-symbols)} and \textit{TIGRE} telescope at La Luz Observatory \textit{(plus-symbols)}. The solid line is the sinusoidal function representing the peak with the highest power in the periodogram (see \ref{sec:period}).}
\label{fig:sindex}
\end{figure}

Our data set comes from different instruments:  Solar-Stellar Spectrograph (\textit{SSS}) at Lowell Observatory in Arizona, \textit{SMARTS} (Small and Moderate Aperture Research Telescope System) $1.5 \ \rm m$ telescope in Chile and \textit{TIGRE} telescope at La Luz Observatory in Mexico. These data are described in the subsequent sections. The chromospheric variability of $\epsilon \ \rm Eri$ was also observed by \textit{HIRES} at the Keck Observatory in Hawaii, and the data were published by \citet{2010ApJ...725..875I}. All {the $S_{\rm MWO}$-index measurements} are shown in \ref{fig:sindex}.

\subsubsection{Lowell Observatory}
The Lowell Observatory Solar-Stellar Spectrograph (\textit{SSS}) records the entire
spectrum surrounding Ca\,II H\&K from $386$ to $401$ nm. Regular observations of
$\sim 100$ stars, including $\epsilon \ \rm Eri$, have been ongoing since 1994. We collected for $\epsilon \ \rm Eri$ a total of $260$ observations from 1994 to the late 2018, and we considered for our analysis the data starting from 2002. We performed the
\textit{SSS} data reduction using a set of IDL routines, employing the {usual sequence
of spectroscopic data reduction steps.}
To obtain the S-index, we first placed each \textit{SSS} H\&K
spectrum on an absolute intensity scale using pseudo-continuum reference
points at $312$ and $400$ nm, as described by  \citet{1995ApJ...438..404H}. We measured
the total residual emission in $0.1$ nm rectangular bandpasses centered on the
H and K line cores, and then convert this raw HK index, $F_{HK}$, to flux and
hence to S-index, following the prescription of \citet{1984A&A...130..353R} with modifications
presented by \citet{2007AJ....133..862H}. The resulting calibration is
quite satisfactory; for the full 25-year SSS time series of 23 flat-activity
stars for which we also have long-term records from Mount Wilson, a linear
regression yields $S_{MWO} = 0.976 S_{SSS} + 0.0044$ (Hall et al. 2019, in prep).
The \textit{SSS} grand mean {$S_{MWO}$-index} we obtained for $\epsilon \ \rm Eri$ is $0.528$. 

\subsubsection{SMARTS}
Observations from the \textit{SMARTS} southern HK project \citep{2009arXiv0909.5464M}
include $148$ low-resolution ($R \sim 2500$) spectra obtained on $74$ distinct epochs between August 
2007 and January 2013 using the \textit{RC Spec} instrument on the 1.5-m telescope at Cerro Tololo 
Interamerican Observatory. Bias and flat-field corrections were applied to the $60$ s integrations, 
and the wavelength was calibrated using standard \textit{IRAF} routines. $S_{MWO}$ values were 
extracted from the reduced spectra following \citet{1991ApJS...76..383D}, placing the instrumental 
measurements onto the Mount Wilson scale using contemporaneous observations from the
\textit{SSS} instrument. {The mean of the $S_{MWO}$ values is $0.513$.} 

\subsubsection{TIGRE telescope}
The \textit{TIGRE} spectra were reduced using version 3 
of the TIGRE/HEROS pipeline \citep{doi:10.1002/asna.201713367},
based on the REDUCE package \citep{2002A&A...385.1095P}.
The pipeline follows the usual steps of reducing echelle CCD frames;
it automatically computes line core indices of activity-sensitive lines,
including the combined Ca\,II H\&K lines, $S_{\rm TIGRE}$.
The computation of the $S_{TIGRE}$ indices is analogous to the computation of the 
{$S_{\rm MWO}$-index and they were converted using the equation $S_{MWO} = 0.0360 + 20.02\cdot S_{TIGRE}$}, according to \citet{2016A&A...591A..89M} (see Fig. 1 of that paper).
The TIGRE data set covers a temporal range from 2013 to the end of 2018, comprising in total $86$ spectra, and{ the mean of these measurements results $S_{MWO}=0.514$}.  

\subsubsection{Chromospheric cycle of $\epsilon \ \rm Eri$}
\label{sec:period}
The $S_{\rm MWO}$-index obtained from the three instruments considered in this work are in good agreement with each other. The short-term scatter of the $S_{\rm MWO}$-index seen in \ref{fig:sindex} is potentially due to the rotational modulation of $\epsilon \ \rm Eri$\footnote{A study of rotational effects is outside the scope of this work. See e.g. \citet{2014A&A...569A..79J} for measurements of rotational modulation of various emission lines of $\epsilon \ \rm Eri$.}.

\citet{2013ApJ...763L..26M} calculated the cycle period based on data {covering the years from 1992 to 2013}. Our time range covers more recent years, until the end of 2018. We consider $S_{\rm MWO}$-index measurements starting in 2002, covering thus the time range of the \textit{XMM-Newton} observations, and we performed the period search on this dataset.

\begin{figure}[!htbp]
\centering
\includegraphics[width=\hsize]{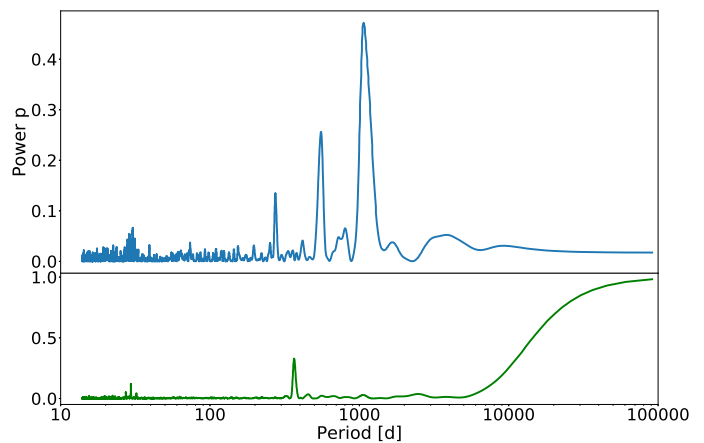}
\caption{Lomb-Scargle periodogram of the Ca\,II S-index data. The bottom panel represents the window function. For the data set that covers in total 15 years, the most significant peak is found at $2.92$ yr.}
\label{fig:lbp}
\end{figure}

The Lomb-Scargle periodogram was calculated using the software \textit{GLS (Generalized Lomb-Scargle Periodogram)}, implemented by \citet{2009A&A...496..577Z}. We found a period of $1067.13$ days. The Lomb-Scargle periodogram is shown in \ref{fig:lbp}, with the window function in the bottom panel. The error on the cycle period was found through Monte-Carlo simulations. We simulated $10000$ data sets of the Ca\,II {$S_{\rm MWO}$-index measurements}. Each data point was randomly drawn from a normal distribution within the observed standard deviation around the measured  {$S_{\rm MWO}$-index}. We performed a Lomb-Scargle analysis for each simulated data set, obtaining $10000$ values of the period; the standard deviation of these values was then considered as the error of the cycle period. To summarize, we found a period of $2.92 \pm 0.02$ yr. This value, and its amplitude resulting from the GLS analysis, were used to plot the sinusoidal function in \ref{fig:sindex}. 

Our result is consistent with the value of $2.95 \pm 0.03$ yr found by \citet{2013ApJ...763L..26M}. However, we did not find the second periodicity at $12.7$ yr. That signal was rooted in the broad minimum from 1984 to 1996, while our data set covers a more recent time range. During the years relevant for our work, i.e. the time covered by \textit{XMM-Newton}, cleary the short cycle was dominant. Moreover, the most recent maximum expected around 2019 was not seen, suggesting another change in cycle behavior. 

\subsection{X-ray data}
Our X-ray monitoring campaign of $\epsilon \ \rm Eri$ started in July 2015. The observations were carried out with the \textit{XMM-Newton} satellite. In this work we present seven snapshots with an exposure time varying between $6$ and $20$ ks.

\label{subsec:specX}
\begin{table}[!htbp]
\centering
\caption{Observing log of \textit{XMM-Newton} data for $\epsilon \ \rm Eri$.}
\begin{tabular}{cccc}
\hline \hline
Date       & Rev. & Science Mode & Exposure Time \\
           &      &   \textit{(EPIC/pn)}           & \textit{(ks)}          \\
\hline
2003-01-19 & 0570 & Full Window  & $11.5$          \\
2015-02-02 & 2775 & Large Window & $16.7$          \\
2015-07-19 & 2858 & Small Window & $6.5$           \\
2016-01-31 & 2957 & Small Window & $7.7$          \\
2016-07-19 & 3042 & Small Window & $8.9$         \\
2017-01-16 & 3133 & Small Window & $6.0$        \\
2017-08-26 & 3244 & Small Window & $8.8$       \\
2018-01-16 & 3316 & Small Window & $6.5$      \\
2018-07-20 & 3408 & Small Window & $19.6$ \\
\hline   
\end{tabular}
\label{tab:obs}
\end{table}

\begin{sidewaystable*}
\caption{Best-fit spectral parameters of each \textit{XMM-Newton} EPIC/pn observation of $\epsilon \ \rm Eri$.}
\resizebox{\textwidth}{!}{%
\begin{tabular}{cccccccccccc}
\hline \hline
Obs No. & Rev  & $kT_1$          & $kT_2$          & $kT_3$          & $logEM_1$        & $logEM_2$        & $logEM_3$        & Flux                            & $L_{\rm X}$ & $T_{\rm av}$ & $\overline{\chi}^2$  \\
&     &                 &                 &                 &                  &                 &                   & $[0.2-2 \ \rm keV]$                   &  $[0.2-2 \ \rm keV]$  & &  \\
&     & $keV$         & $keV$         & $keV$         & $cm^{-3}$        & $cm^{-3}$        & $cm^{-3}$        & $10^{-11} erg \ cm^{-2} \ s^{-1}$ & $10^{28} erg \ s^{-1}$ & $keV$  &\\ 
\hline
1 & 0507 & $0.15 \pm 0.01$ & $0.33 \pm 0.02$ & $0.75 \pm 0.02$ & $50.70 \pm 0.06$ & $50.83 \pm 0.04$ & $50.55 \pm 0.04$ & $1.29 \pm 0.01$  & $1.58 \pm 0.01$ & $0.37 \pm 0.02$ & $1.02$\\
2 & 2775 & $0.15 \pm 0.02$ & $0.33 \pm 0.03$ & $0.78 \pm 0.05$ & $50.65 \pm 0.09$ & $50.81 \pm 0.06$ & $50.39 \pm 0.08$ & $1.10 \pm 0.03$  & $1.35 \pm 0.04$ & $0.35 \pm 0.03$ & $0.81$\\
3 & 2858 & $0.15 \pm 0.02$ & $0.34 \pm 0.02$ & $0.77 \pm 0.04$ & $50.81 \pm 0.05$ & $51.01 \pm 0.04$ & $50.60 \pm 0.05$ & $1.74 \pm 0.02$  & $2.13 \pm 0.02$ & $0.37 \pm 0.02$ & $1.08$\\
4 & 2957 & $0.16 \pm 0.02$ & $0.33 \pm 0.02$ & $0.74 \pm 0.04$ & $50.78 \pm 0.08$ & $51.00 \pm 0.04$ & $50.57 \pm 0.06$ & $1.69 \pm 0.02$  & $2.06 \pm 0.02$ & $0.35 \pm 0.02$ & $1.00$\\
5 & 3042 & $0.19 \pm 0.02$ & $0.36 \pm 0.03$ & $0.79 \pm 0.03$ & $50.85 \pm 0.08$ & $50.87 \pm 0.10$ & $50.83 \pm 0.04$ & $1.92 \pm 0.02$  & $2.35 \pm 0.02$ & $0.44 \pm 0.03$ & $1.00$ \\
6 & 3133 & $0.19 \pm 0.03$ & $0.33 \pm 0.04$ & $0.73 \pm 0.03$ & $50.73 \pm 0.12$ & $50.83 \pm 0.13$ & $50.64 \pm 0.05$ & $1.47 \pm 0.02$  & $1.80 \pm 0.02$ & $0.39 \pm 0.05$ & $0.95$\\
7 & 3244 & $0.15 \pm 0.02$ & $0.32 \pm 0.02$ & $0.72 \pm 0.04$ & $50.68 \pm 0.08$ & $50.95 \pm 0.04$ & $50.47 \pm 0.06$ & $1.40 \pm 0.01$  & $1.72 \pm 0.02$ & $0.34 \pm 0.02$ & $1.17$\\
8 & 3316 & $0.21 \pm 0.02$ & $0.36 \pm 0.08$ & $0.78 \pm 0.03$ & $50.89 \pm 0.17$ & $50.85 \pm 0.20$ & $50.83 \pm 0.05$ & $1.97 \pm 0.02$  & $2.41 \pm 0.02$ & $0.43 \pm 0.05$ & $0.90$\\
9 & 3408 & $0.17 \pm 0.01$ & $0.34 \pm 0.01$ & $0.79 \pm 0.01$ & $50.70 \pm 0.05$ & $51.03 \pm 0.04$ & $50.85 \pm 0.02$ & $2.08 \pm 0.01$  & $2.55 \pm 0.01$ & $0.44 \pm 0.02$ & $1.60$\\
\hline
\end{tabular}
}
\label{tab:spec}
\end{sidewaystable*}

\begin{figure*}[!htbp]
\centering
\includegraphics[width=\hsize]{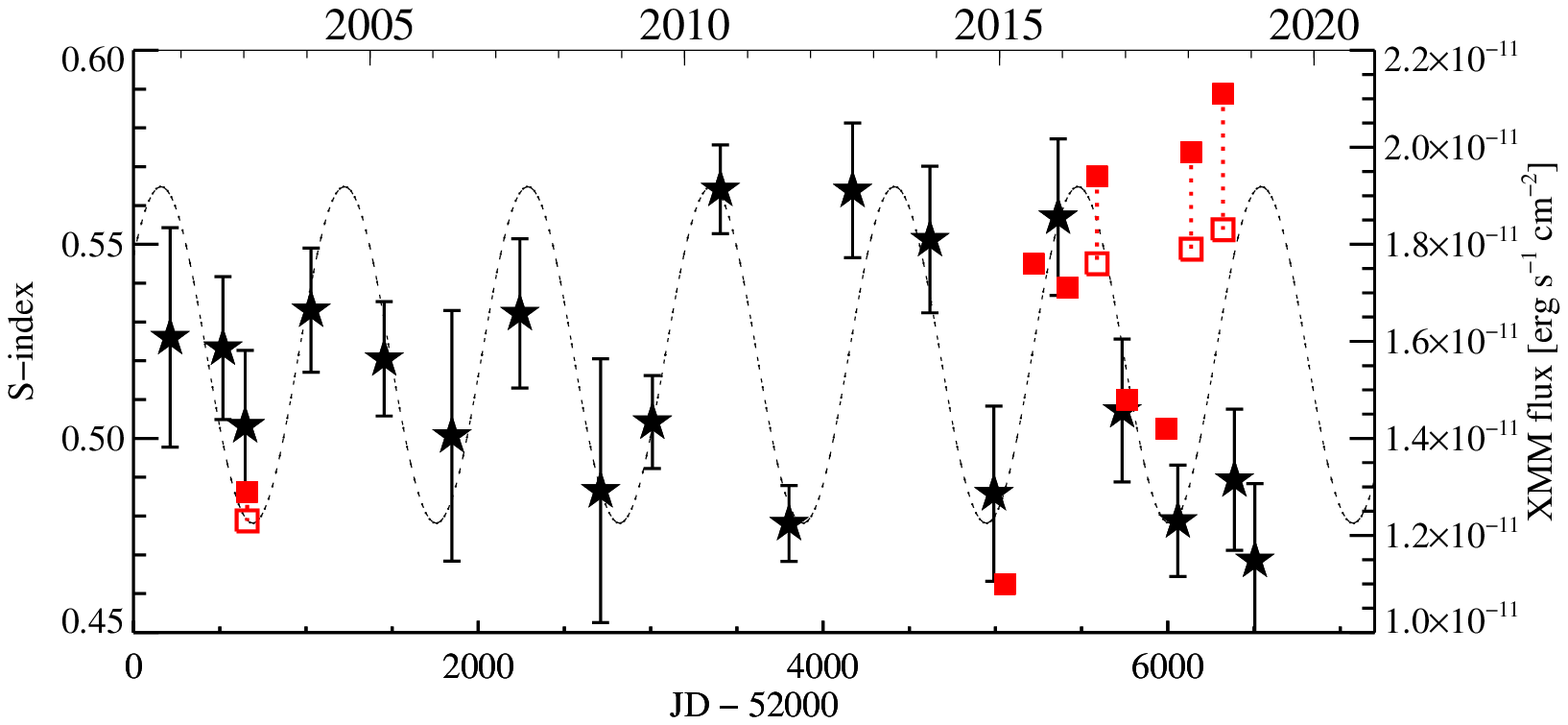}
\caption{Long-term lightcurve of $\epsilon \ \rm Eri$ starting from 2002. The binned Ca\,II S-index data are plotted with star symbols. The X-ray fluxes are overplotted as red squares. The filled squares refer to the X-ray fluxes calculated over the complete lightcurve of each observation. For those observations with a flare-like short term variability, the X-ray fluxes are also calculated after the removal of the flare and shown as open squares.}
\label{longterm_lc}
\end{figure*}
\textit{XMM-Newton} allows to monitor the target using two X-ray instruments simultaneously: EPIC (European Imaging Photon Camera) pn detector covers the energy band $0.15-15 \ \rm keV$ and EPIC MOS detector $0.2-10 \ \rm keV$;
RGS (Reflection Grating Spectrometer) produces high-resolution spectra. Our observations were conducted in EPIC Small Window mode. Previous to our campaign, $\epsilon \ \rm Eri$ had been observed twice by \textit{XMM-Newton} in Full and Large Frame mode. These two observations are also considered in our analysis. The observing log is given in \ref{tab:obs}. {We focus on the analysis of the data from the EPIC/pn instrument: adding the EPIC/MOS would considerably complicate the subsequent analysis (see \ref{sec:sim}), without adding relevant information since its signal-to-noise ratio is lower than that of the EPIC/pn.}

The X-ray data were analyzed with the software SAS (Science Analysis System - version 17.0.0). The standard SAS tools, as described in the SAS Users Guide \citep{SAS}, were applied to filter event lists of each observation and produce the images.

\subsubsection{EPIC spectra}

We extracted the spectra from the filtered EPIC/pn event lists. We first chose two circular regions to extract the source and the background counts. The source region was centered on the detected source, and the background region on a source-free section of the CCD. We then extracted the spectra from these regions and we scaled them to the chosen area. We generated the redistribution matrix and the ancillary files for each source spectrum, and grouped these together with the spectrum file choosing $40$ as minimum number of counts per bin.  

Before analyzing the spectra, we corrected them for the so called Out-of-Time events, i.e. EPIC camera photon events registered during the readout of the CCD.
In the Full Frame mode and in the Large Frame mode the Out-of-time events are $6.3 \%$ and $0.16 \%$ of the total registered photons, while in the Small Window mode they are $1.1 \%$ \footnote{The percentage of Out-of-Time events for the different observational EPIC modes is reported in the \textit{\textit{XMM-Newton} Users Handbook}.}. Thus, using SAS we extracted first the spectra of the Out-of-Time events for each observation and then we subtracted them from the respective source spectrum.  

All spectra were analyzed with the software \textit{xspec} (version 12.10; \citealt{1996ASPC..101...17A}). We first considered the merged spectrum, including all observations: we fitted it assuming a $\text{3-T}$ APEC model where the temperatures, the emission measures and the global abundance are allowed to vary. We did not include an absorption component in our model because, due to the small distance of $\epsilon \ \rm Eri$, photoeletric absorption is negligible. We obtained as best-fitting parameter of the abundance $0.29 Z_{\odot}$. Similarly, when we fit each single spectrum with the same model, the abundances span a range between $0.2 Z_{\odot}$ and $0.4 Z_{\odot}$. While these values are typical for very young stars \citep{2007ApJ...660.1462M}, from the high-resolution spectrum of $\epsilon \ \rm Eri$
slightly larger abundances ($[Fe]/[H]=0.5 \pm 0.2$) were found \citep{2004A&A...416..281S}\footnote{We note that all these values refer to the {\it coronal} abundances
which are measured with respect to the solar {\it photospheric} abundance.}.
A detailed abundance study is beyond the scope of this work. Thus, to minimize the number of free parameters and simplify the subsequent analysis described in \ref{sec:sim}, we decided to set the metal abundances on $0.3 Z_{\odot}$ and to keep them frozen during the fitting procedure.  

The best-fit spectral parameters of each observation are summarized in \ref{tab:spec}. The associated errors were found with the \textit{xspec} command \textit{error}, that determines the confidence interval for the model parameters within a confidence region that we chose equal to $90 \%$. We calculated the fluxes in the soft energy band $0.2-2$~keV, the luminosities $L_X$ and the emission-measure weighted average temperatures as 
\begin{equation}
T_{\rm av} = \dfrac{\sum _{i=1} ^3 kT_i \cdot EM_{i}}{\sum _{i=1} ^3 EM_{i}}.
\label{eq:1}
\end{equation}

\ref{longterm_lc} shows the resulting long-term X-ray lightcurve together with the Ca\,II S-index and the sinusoidal fit resulting from the Lomb-Scargle analysis of the S-index. Here, we  binned the Ca\,II measurements of \ref{fig:sindex} for clarity\footnote{We divided the $S_{\rm MWO}$-index set in sub-sets, each of them with a length corresponding to an observing season; the vertical bars are the standard deviation of each sub-set.}. 

Clearly, the X-ray fluxes in \ref{longterm_lc} follow the sinusoidal fit to the Ca\,II variability until late 2017, providing evidence of an X-ray activity cycle. However, the two observations of 2018 show an enhanced X-ray flux. In the next section we show that these observations are affected by short-term variability.

\subsubsection{EPIC lightcurves}
\label{sec:lc}

EPIC/pn lightcurves were generated in the soft energy band $0.2-2$~keV for a time bin size of $300$ sec. 

We systematically analyzed the lightcurves of each individual observation for variability, using the software \textit{R} and its package \textit{changepoint} \citep{JSSv058i03}. A \textit{changepoint} is denoted as the time at which a significant change of the count rate is present. This tool allows us to identify multiple changes in mean and variance of the count rates within each observation. In Appendix A the lightcurves of all observations and their segmentation are shown. According to this analysis, four out of nine observations show short-term variability. These observations are the ones from January 2003, July 2016, January 2018 and August 2018. The lightcurves of January 2003 and July 2016 show a sudden increase in the count rate resembling the shape of a stellar flare event: in particular the one of 2003 shows most likely the decay phase of a flare. We thus decided for these observations to extract the spectra in the segment of the lightcurve with the lowest count rate and repeat the spectral fitting. The lower X-ray fluxes resulting from this analysis are also plotted in \ref{longterm_lc} as open red squares.

\section{The corona of $\epsilon \ \rm Eri$ in the context of solar EMD}
\label{sec:sim}
Here we aim at describing the \textit{XMM-Newton}/EPIC spectrum of $\epsilon \ \rm Eri$ and its evolution throughout its X-ray cycle in terms of solar emission measure distributions (EMDs).

In the context of the study of the ``Sun as an X-ray star'' \citep{2000ApJ...528..524O, 2000ApJ...528..537P, 2001ApJ...557..906R,2001ApJ...560..499O}, \citet{2000ApJ...528..524O} generated a grid of the temperatures $T$ and the emission measures $EM$ of individual pixels from \textit{Yohkoh}/SXT images of the Sun taken in two different filters during the 1990's. Combining the results over the full surface of the Sun, a ``whole-Sun'' $EM(T)$ distribution (EMD) was constructed. This analysis was extended to studies of both spatially and temporally distinct observations of the Sun in subsequent articles of the paper series, that here are briefly summarized.

While the analysis of  \citet{2000ApJ...528..524O} was restricted to a single observing date (namely January 6th, 1992), \citet{2000ApJ...528..537P} have extended the study to three different dates, spanning the full range of the solar activity cycle. That work showed that the solar EMD changes strongly throughout the cycle.  
\citet{2001ApJ...557..906R} derived the EMDs of flaring events (FLs): they chose eight flares, from weak to intense ones (i.e. from class C to class X), as representative of the flaring Sun, and 
they analyzed separately their rise, peak and decay phase.
\citet{2001ApJ...560..499O} defined three types of solar coronal structures based on the intensity measured in the \textit{Yohkoh}/SXT images. The full range of measured intensities was assigned to quiet regions (the background corona, BKCs), active regions (ARs) and cores of active regions (COs) in order of increasing intensity. This classification was verified to closely correspond to the distinct spatial structures seen in the images. The EMDs were constructed for each of these type of structures separately during the whole solar cycle. It was, thus, noticed that during the minimum of the cycle the dominant contribution to the EMD comes from the BKCs, while during the maximum from the ARs.
Finally, \citet{2004A&A...424..677O} examined the temporal evolution of the EMDs of only one visible AR and one CO on the solar surface in time-steps of one day and spanning \textit{Yohkoh}/SXT observations across nearly two months. 

The various solar EMDs obtained from these studies were converted into synthetic X-ray spectra using the MEKAL code \citep{1986A&AS...65..511M, 1992K, 1995Leg..16M}, as described by \citet{2000ApJ...528..537P}. These spectra can be folded through the instrumental response of non-solar instruments, such as \textit{ROSAT}, \textit{ASCA}, \textit{XMM-Newton}. The final products of the 
``Sun as an X-ray star'' studies were simulated X-ray spectra of the Sun that are analogous to stellar observations and that can be treated with the usual methods of X-ray analysis of the chosen non-solar instrument. 

\citet{2008A&A...490.1121F} and \citet{2017A&A...605A..19O} had applied this study to the \textit{XMM-Newton} spectra of the star HD 81809, with the aim of using the Sun as a template to link the stellar coronae physics to the standard solar model.
They combined the solar EMDs extracted for ARs, COs and FLs and scaled them to the size of HD 81809. By varying the coverage fractions of each solar structure on the surface of the star, they had then built a grid of EMDs to artificially reproduce a solar-like corona with the physical characteristics of HD 81809. Subsequently, they extracted X-ray spectra from the grid to be compared with the observed EPIC spectra of HD 81809 to interpret the evolution of the X-ray cycle in terms of coronal structures. This can only be achieved by a spectral analysis since, in contrast to the Sun, the magnetic structures on the artificial solar-like corona are not spatially and temporally resolved and diagnosed.
Here, we apply the same method to $\epsilon \ \rm Eri$.

\subsection{Standard solar coronal structures}
\label{subsec:std_emd}

\begin{figure*}[!htbp]
\centering
\subfloat{\includegraphics[width=\columnwidth]{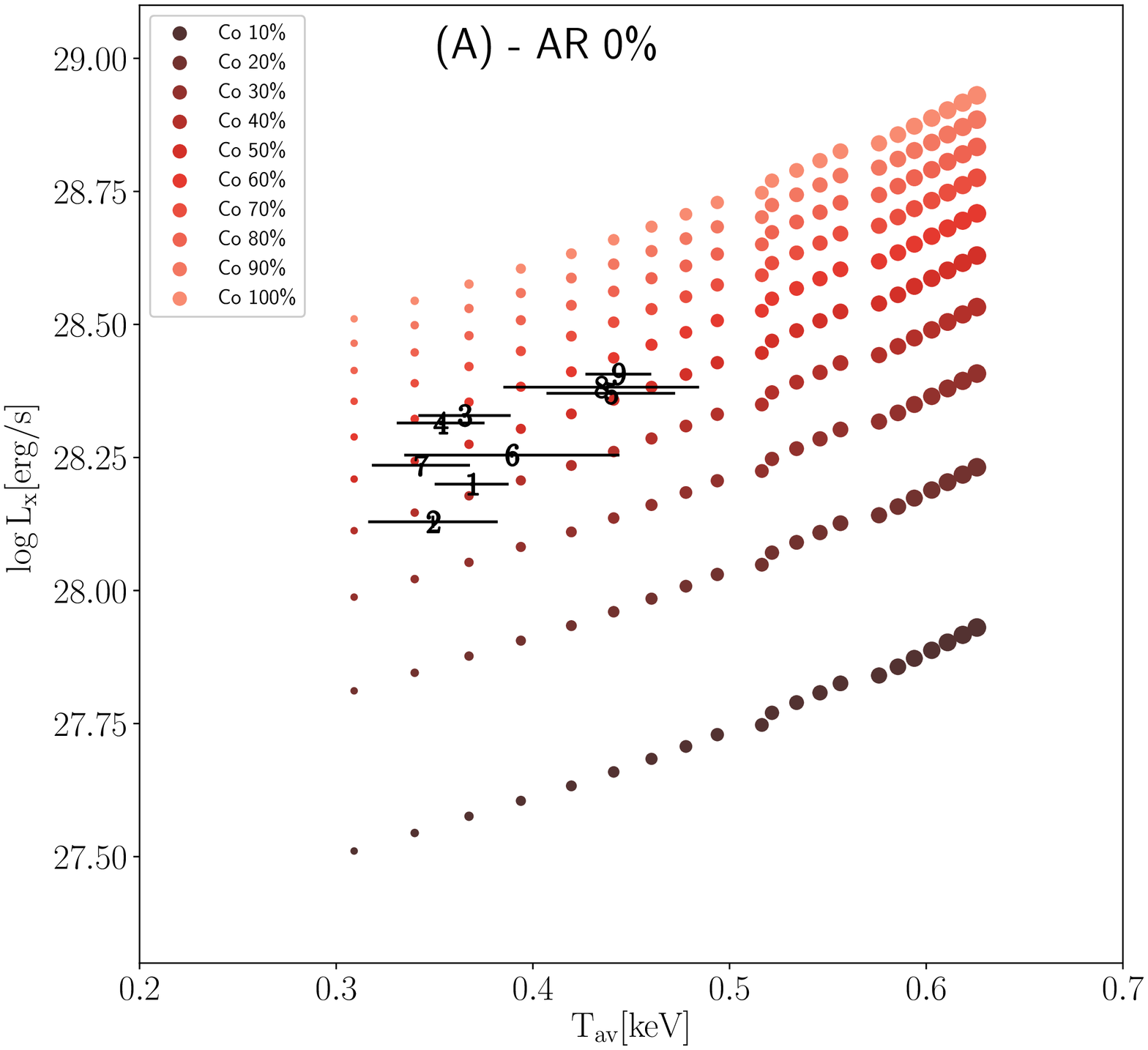}} 
\subfloat{\includegraphics[width=\columnwidth]{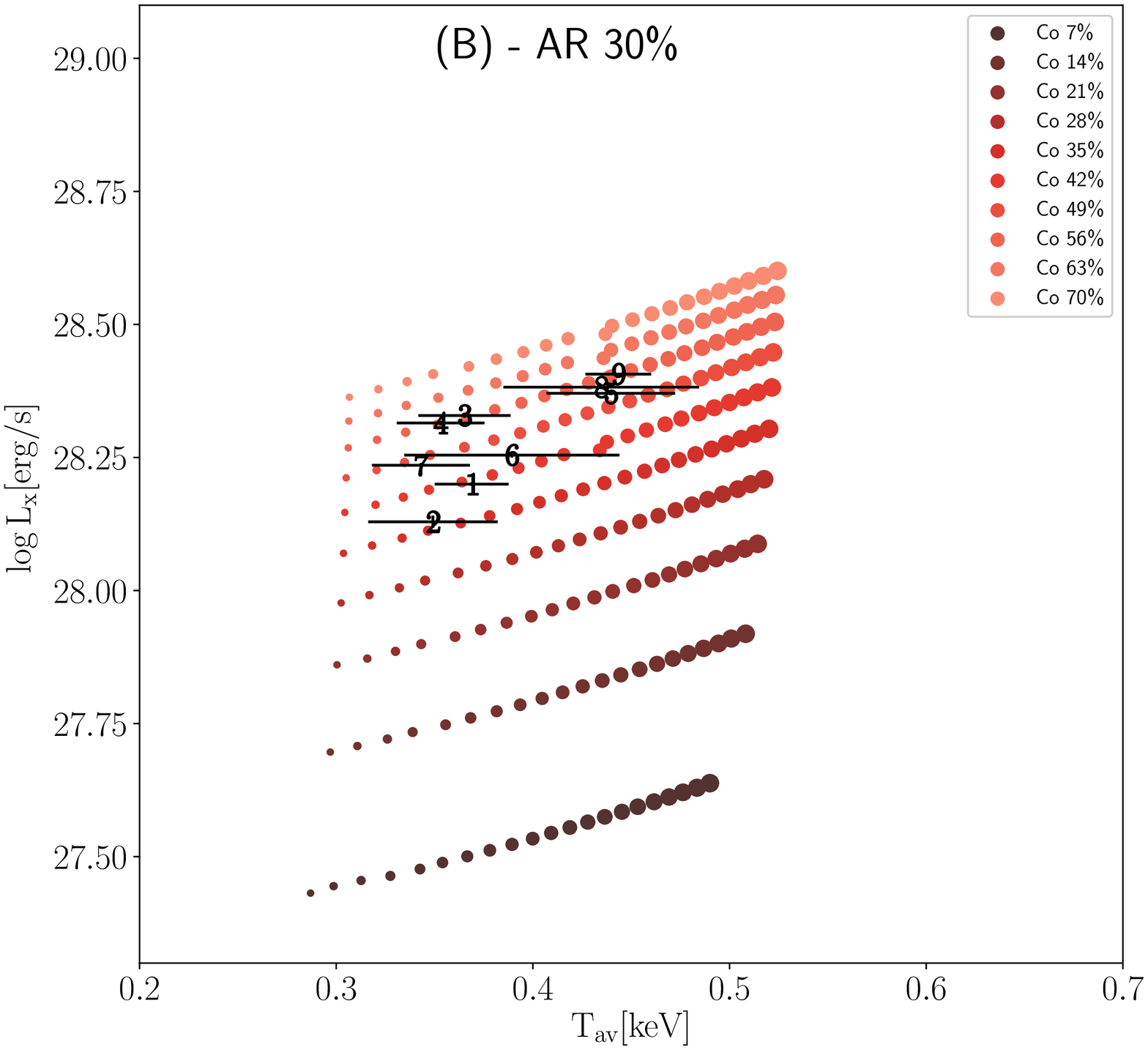}} \\
\subfloat{\includegraphics[width=\columnwidth]{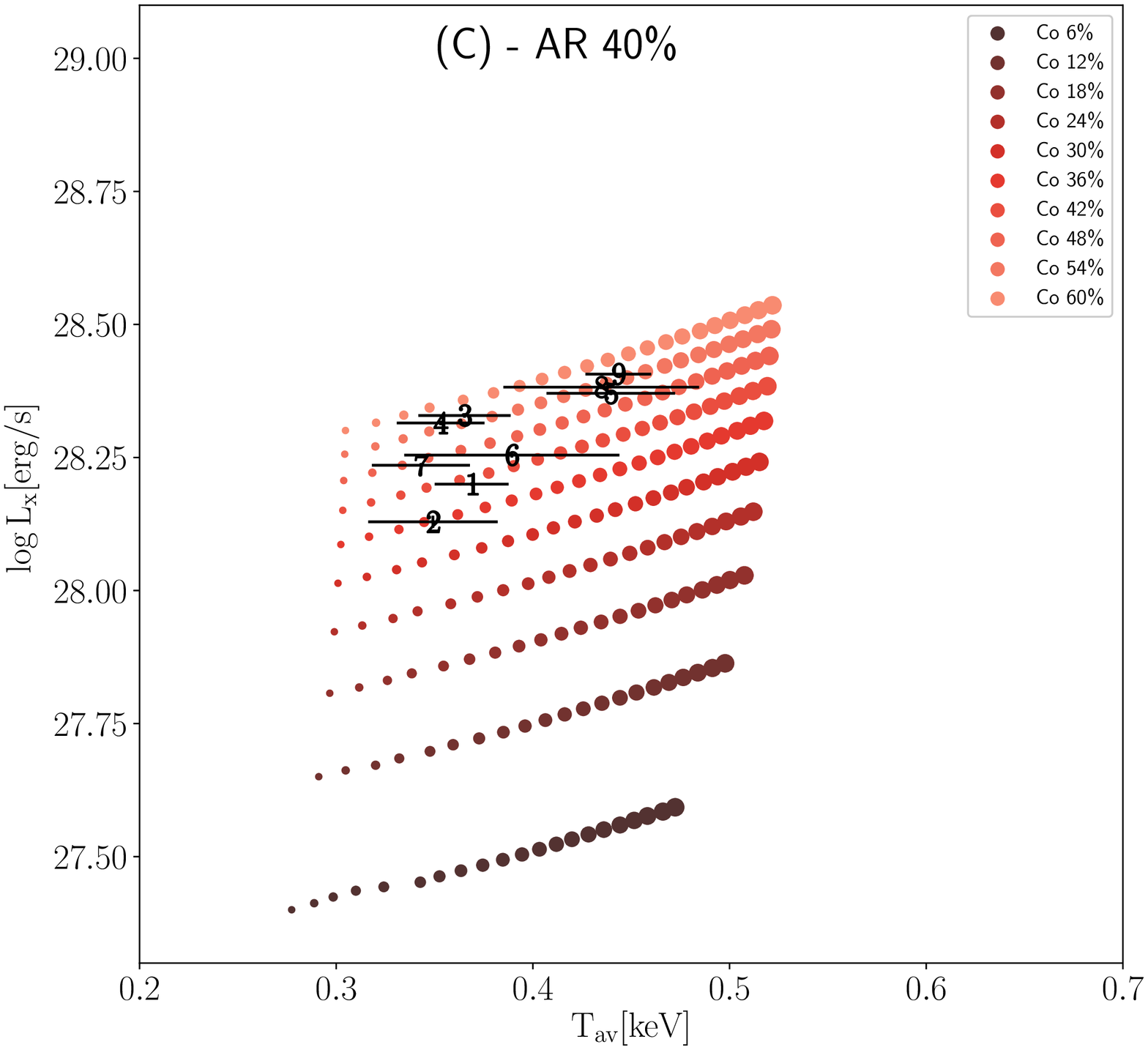}} 
\subfloat{\includegraphics[width=\columnwidth]{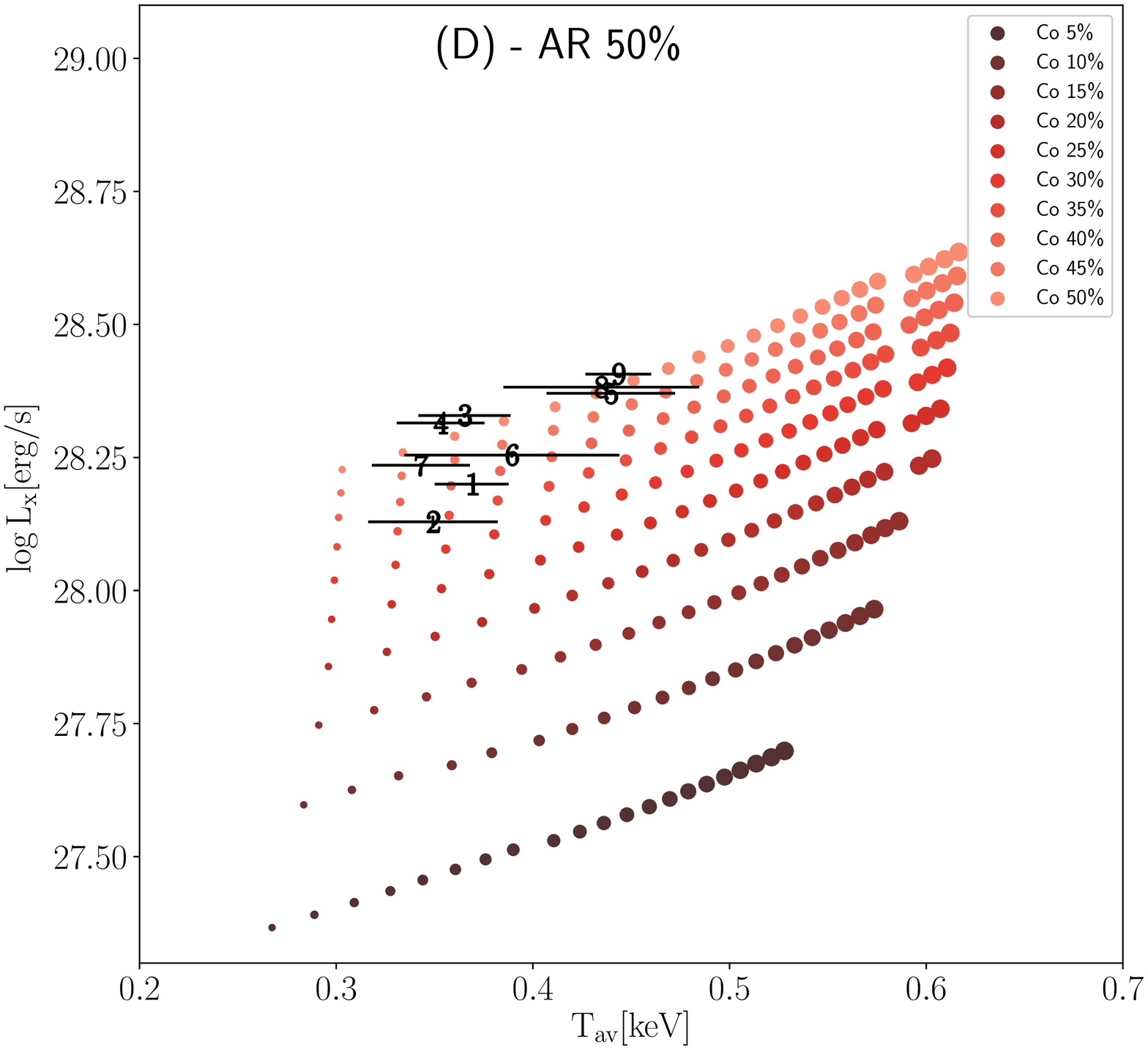}} \\
\caption{Four combinations of standard coronal structures explored: (a) AR equal to $0 \%$ and COs varying between $10 \%$ and $100 \%$; (b) AR equal to $30 \%$ and COs varying between $7 \%$ and $70 \%$; (c) AR equal to $40 \%$ and COs varying between $6 \%$ and $60 \%$; (d) AR equal to $50 \%$ and COs varying between $5 \%$ and $50 \%$. For all the combinations the FLs are allowed to vary between $0 \%$ and $2 \%$ of the CO area fraction. Increasing CO coverage is represented by darker color, while increasing FL fraction is represented by increasing symbol size. The overplotted numbers are the observed values of $\epsilon \ \rm Eri$ and are chronologically ordered.}
\label{fig:ar0}
\end{figure*}
Among the coronal structures described and analyzed in the study of the ``Sun as an X-ray star'', we used ARs, COs and FLs. The emission measure of $\epsilon \ \rm Eri$ results to be higher than the solar one. Thus, we ignored the BKCs because, among all types of structures, they have the lowest intensity and a high percentage of coverage would be required to produce a significant contribution to the total EMD. This would considerably reduce the available surface for the other magnetic structures, that instead, given their higher intensity, are better suited to reproduce the $L_{\rm X}$ of $\epsilon \ \rm Eri$. Moreover, the solar BKCs have low temperatures that are only marginally covered with \textit{XMM-Newton}. \citet{2000ApJ...545.1074D} showed that the EMD of $\epsilon \ \rm Eri$ at temperatures representative of solar BKC ($T \sim 10^6$ K) obtained from \textit{Extreme Ultraviolet Explorer (EUVE)} observations is significantly greater than that of the full Sun. In \ref{sec:disc} we will show that the EMD of $\epsilon \ \rm Eri$ derived with \textit{XMM-Newton} at these temperatures well matches the one obtained from the EUVE.

For the $EM(T)$ distribution of ARs and COs, we took into account the time-averaged distribution of these structures presented by \citet{2004A&A...424..677O}, i.e. the average of the evolution of only one solar active region and one core observed by \textit{\textit{Yohkoh}} in 1996, from its emergence to its decay. For the flare $EM(T)$ we considered the time-averaged distribution of the eight flares discussed in \citet{2001ApJ...557..906R}. The assumptions to derive the flaring contributions are the same given in \citet{2017A&A...605A..19O},  which take into account the differential flare energy distribution $N(E) \propto E^{\alpha} dE$ of the Sun with index $\alpha=1.53$ (see \citealt{2017A&A...605A..19O} for details).

We produced the EMDs for unit surface area for the metal abundance used in the fits of the EPIC/pn spectra of  $\epsilon \ \rm Eri$, i.e. $0.3Z_{\odot}$. However, the EMDs of each solar structure had been extracted considering solar metal abundances. We thus scaled the EMDs by a factor of $1/0.3$ that compensates for the reduced radiative losses associated with lower abundances. 
Then we obtained the EMDs of each magnetic structure by multiplying the EMDs per unit surface area by the stellar area of $\epsilon$~Eri that might be covered by these structures.

\begin{figure}
\centering
\includegraphics[width=\columnwidth]{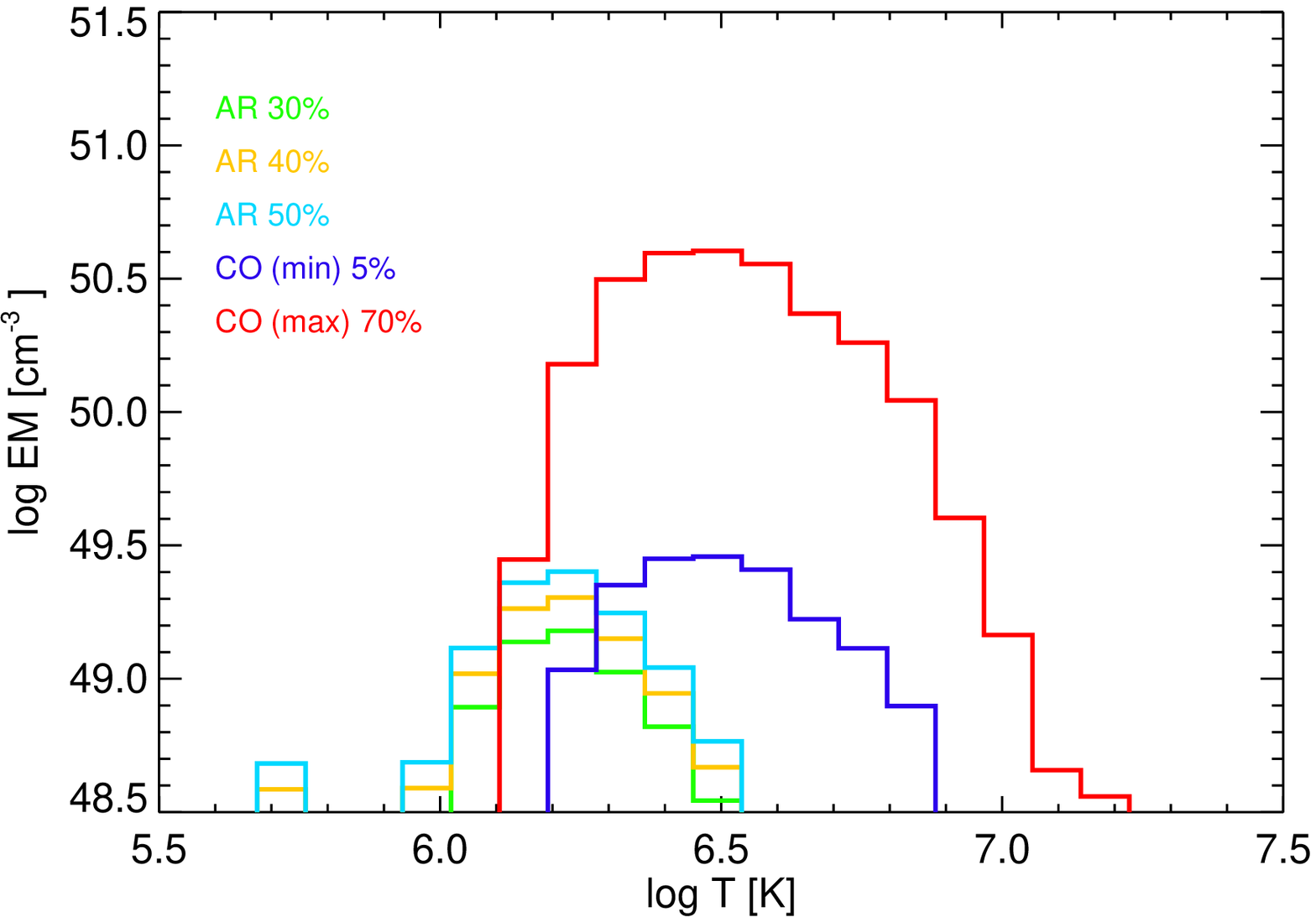}
\caption{EMDs of the three combinations of ARs shown in \ref{fig:ar0}b-d scaled to the surface area of $\epsilon \ \rm Eri$. The EMD of ARs are plotted in green, in yellow and in cyan for each investigated coverage fraction. The EMD of COs are also overplotted in blue for the minimum of the considered coverage fraction and in red for the maximum.}
\label{fig:emd_ARCO}
\end{figure}

\begin{figure*}[]
\centering
\subfloat{\includegraphics[width=\columnwidth]{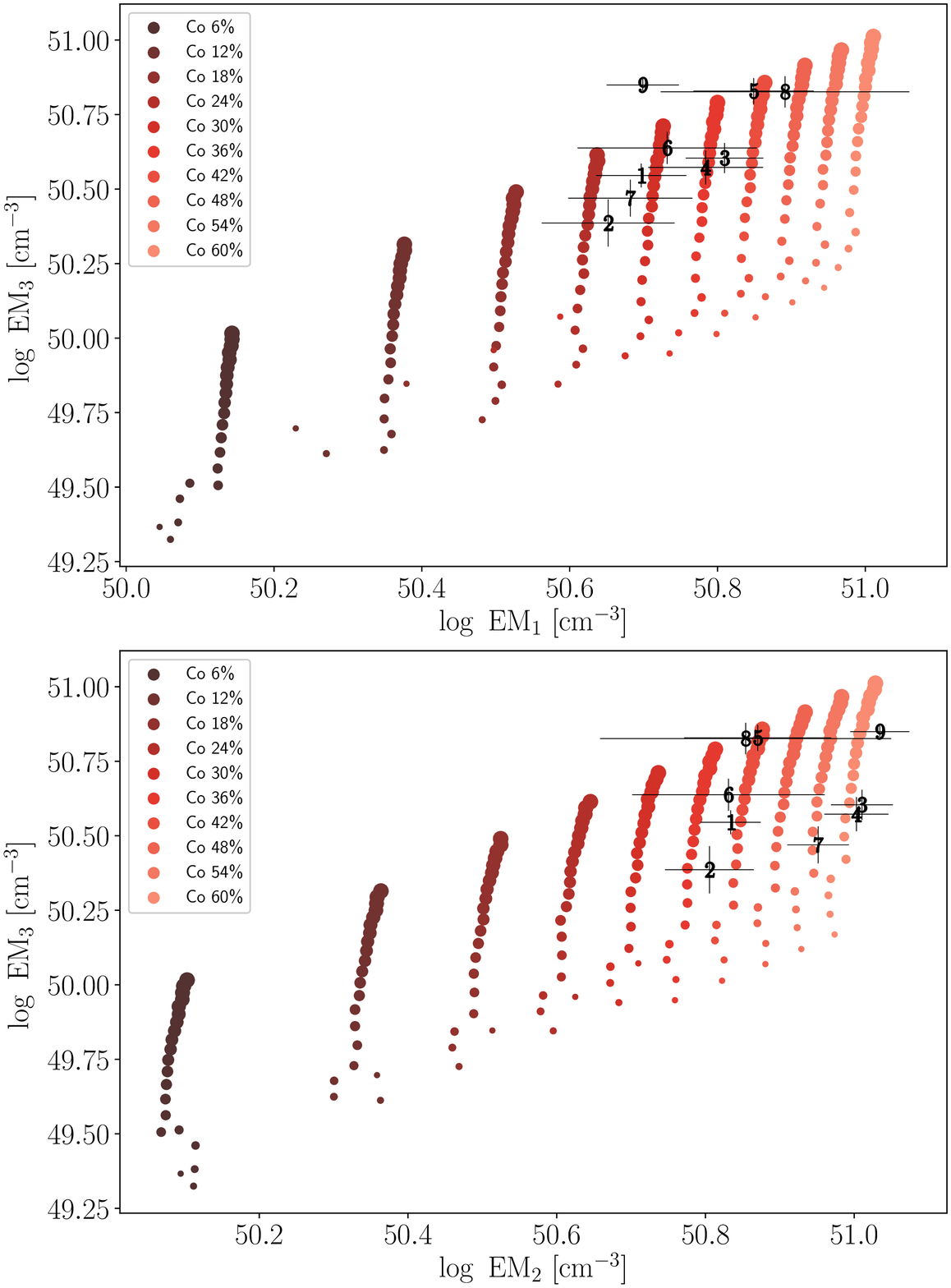}} 
\subfloat{\includegraphics[width=\columnwidth]{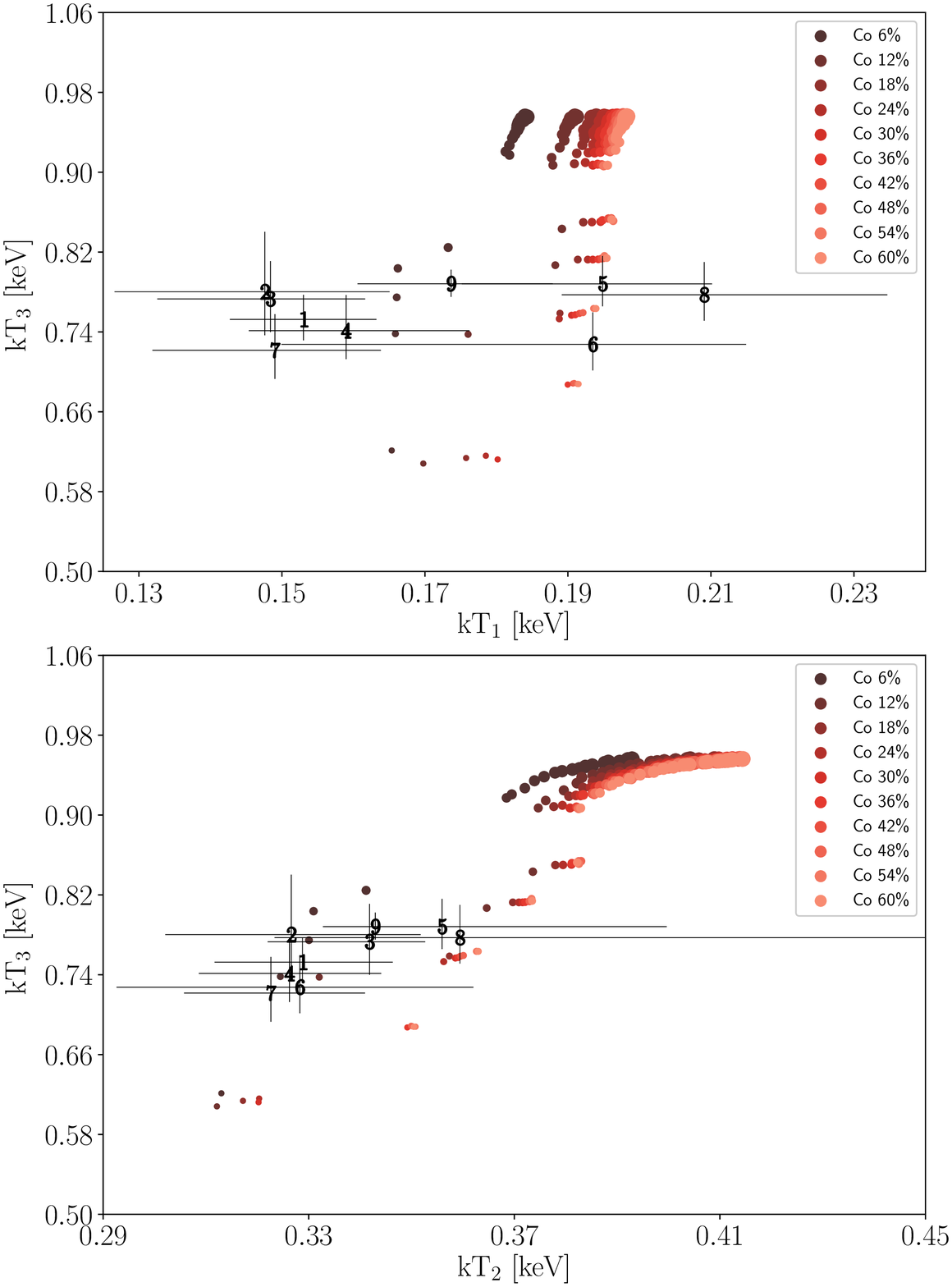}} \\
\caption{Parameters of the best-fittting 3T-model for the individual observations of $\epsilon \ \rm Eri$  (numbers) with overplotted the best-fit parameters obtained from the synthetic spectra for the grid from \ref{fig:ar0}c. On the left, the emission measures. On the right, the temperatures. Colors and symbol sizes are the same as in \ref{fig:ar0}.}
\label{fig:ar40_par}
\end{figure*}
\begin{figure}[!htbp]
\includegraphics[width=\columnwidth]{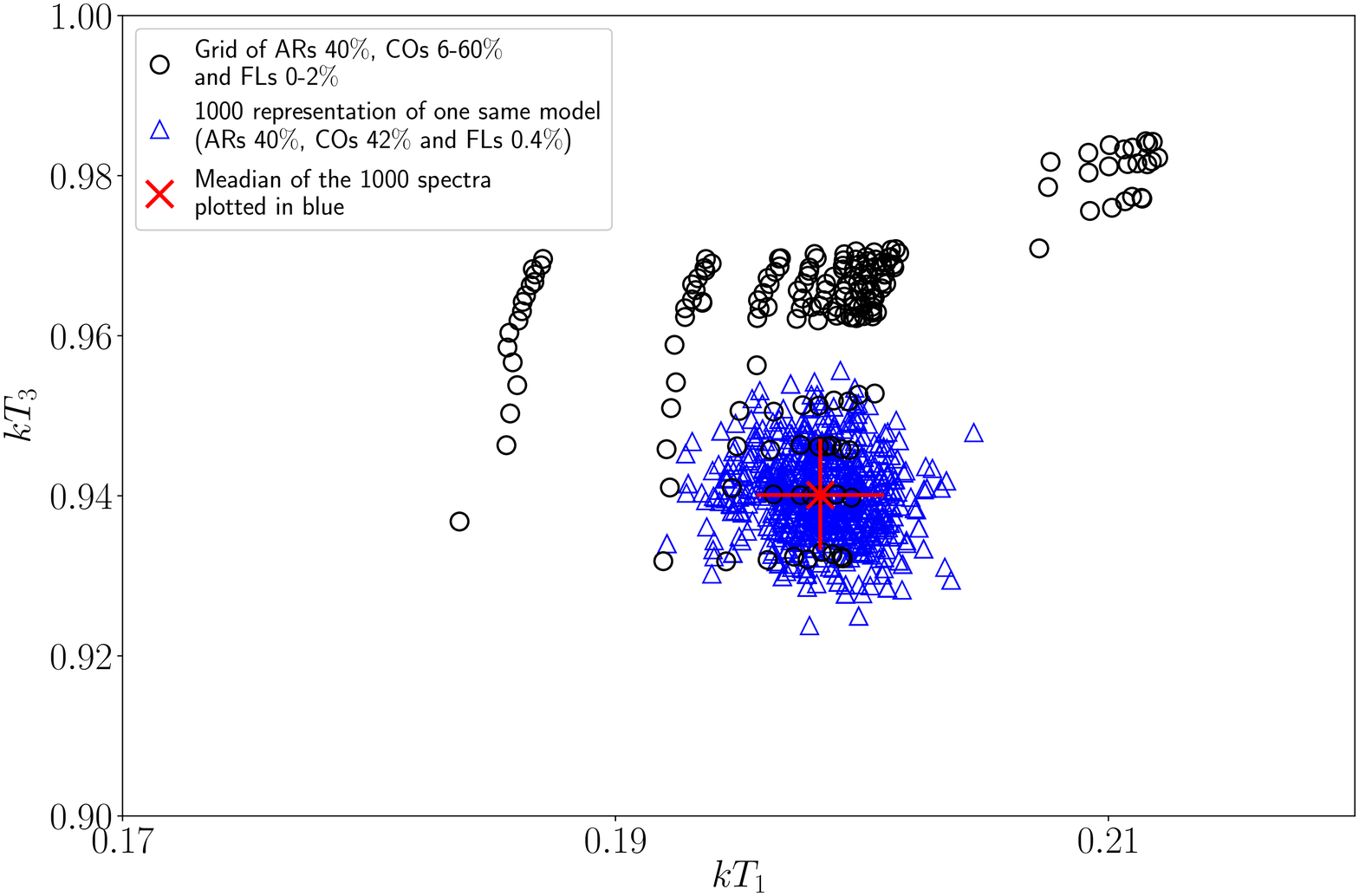}
\caption{Example showing the effect of the Poisson statistics on the retrieved spectral parameters ($kT_1$ vs. $kT_3$) of the synthetic spectra of the grid with ARs $40\%$, COs $6-60\%$ and FLs $0-2\%$. \textit{Black circles} - medians of the 1000 representations of each grid point; \textit{Blue triangles} - 1000 best fit parameters representing one specific combination of magnetic structures (ARs $40\%$, COs $42\%$, FLs $0.4\%$); \textit{Red cross} - median of the 1000 representations (plotted in blue) of the model with COs $42\%$ and FLs $0.4\%$ that we chose to highlight as an example.}
\label{fig:spread}
\end{figure}

The task is now to determine the relative contribution of ARs, COs and FLs to the X-ray emission of $\epsilon \ \rm Eri$. We constructed thus a grid of $EM(T)$ distributions where each grid point represents a different percent coverage of the coronal structures on the surface of $\epsilon \ \rm Eri$.
We synthesized the EPIC/pn spectra for each grid point of the composite $EM(T)$ characterized by different fractional contributions of the three magnetic structures and assuming abundances $0.3Z_{\odot}$. 
We analyzed each synthetic spectrum in the same way as we had done for the observed spectra of $\epsilon \ \rm Eri$, i.e. we fitted them with a 3-T thermal APEC model with metal abundance fixed at $0.3 \ Z_{\odot}$ (\ref{subsec:specX}).
We then calculated for each grid point the X-ray luminosities ($L_{\rm X}$) and the average temperatures ($T_{\rm av}$) as in \ref{subsec:specX}. 

In \ref{fig:ar0}a-d we explore how different combinations of ARs, COs and FLs affect the derived $L_{\rm X}$ and $T_{\rm av}$. In these panels the results for the synthetic spectra are represented by dots. The $L_{\rm X}$ and the $T_{\rm av}$ derived for $\epsilon \ \rm Eri$ are overplotted onto this grid with numerical symbols, following the chronological order of each observation. In \ref{fig:ar0}a the AR coverage fraction is set to $0 \%$, while the COs are varying between $10 \%$ and $100 \%$ as shown in the inset. The FLs vary between $0 \%$ and $2 \%$ of the area covered by the COs, with a step size of $0.1 \%$. The symbol size and the color in the plots represent, respectively, the variation of the percentage of FLs and COs. It can be seen that an increase in the coverage fraction of the FLs influences mainly the average temperature, whereas an increase in the coverage fraction of the COs influences the luminosity.  We notice that the observed $L_{\rm X}$ and $T_{\rm av}$ of $\epsilon \ \rm Eri$ are well reproduced with a surface coverage with COs that does not exceed $60 \%$. 

The impact of adding AR coverage greater than $0 \%$ can be seen in panels b-d. Obviously, the sum of the coverage fraction of ARs and COs can not exceed $100 \%$.
To set an upper limit for the contribution of ARs and COs, we explored three different combinations of coronal structures around the percentage of the COs of $60 \%$. We chose  ARs equal to $30 \%$ and COs varying between $7 \%$ and $70 \%$  (\ref{fig:ar0}b); ARs equal to $40 \%$ and COs varying between $6 \%$ and $60 \%$  (\ref{fig:ar0}c); ARs equal to $50 \%$ and COs varying between $5 \%$ and $50 \%$  (\ref{fig:ar0}d). For all the combinations shown in \ref{fig:ar0}, the flares are allowed to vary between $0 \%$ and $2 \%$ of the percentage of COs. In \ref{fig:emd_ARCO} the EMDs of the ARs for each combination are shown, together with the minimum and the maximum of COs that we considered in all the combinations.

As can be seen from \ref{fig:ar0}b-d, all three combinations can potentially reproduce our observational data. We reject the combination of \ref{fig:ar0}d because the $L_{\rm X}$ of the synthetic spectra cover the data only at the limit. The other combinations reproduce the observations. However, we can set only an upper limit for the coverage fraction of ARs. As matter of fact, the emission measure of the ARs is lower than that of the COs per unit of covered surface, as is evident in \ref{fig:emd_ARCO} where the EMD of the minimum coverage fraction of COs, i.e. $5 \%$, has an emission measure comparable to the maximum coverage fraction of ARs, i.e. $50 \%$. Thus, any reasonable coverage fraction of ARs provides only a small contribution to the overall EMD, and the difference between this contribution for different AR percentage is small (see \ref{fig:emd_ARCO} for $30 \%$, $40\%$ and $50\%$ of AR) such that we can not set a lower limit on the coverage of ARs. Thus, the grid chosen for further analysis is the combination with a coverage of ARs equal to $40\%$,
a coverage of COs that varies from $6\%$ to $60\%$ and FLs varying from $0\%$ to $2\%$ (\ref{fig:ar0}c), without excluding that the surface can also be covered with a lower fraction of ARs. 

For a more detailed investigation of the compatibility between the observed spectra of $\epsilon \ \rm Eri$ and the synthetic spectra derived  from the solar $EM(T)$, we proceeded to a comparison of the individual spectral best fit parameters, i.e. the three temperatures and the three emission measures. 

\begin{figure*}
\centering
\includegraphics[width=\hsize]{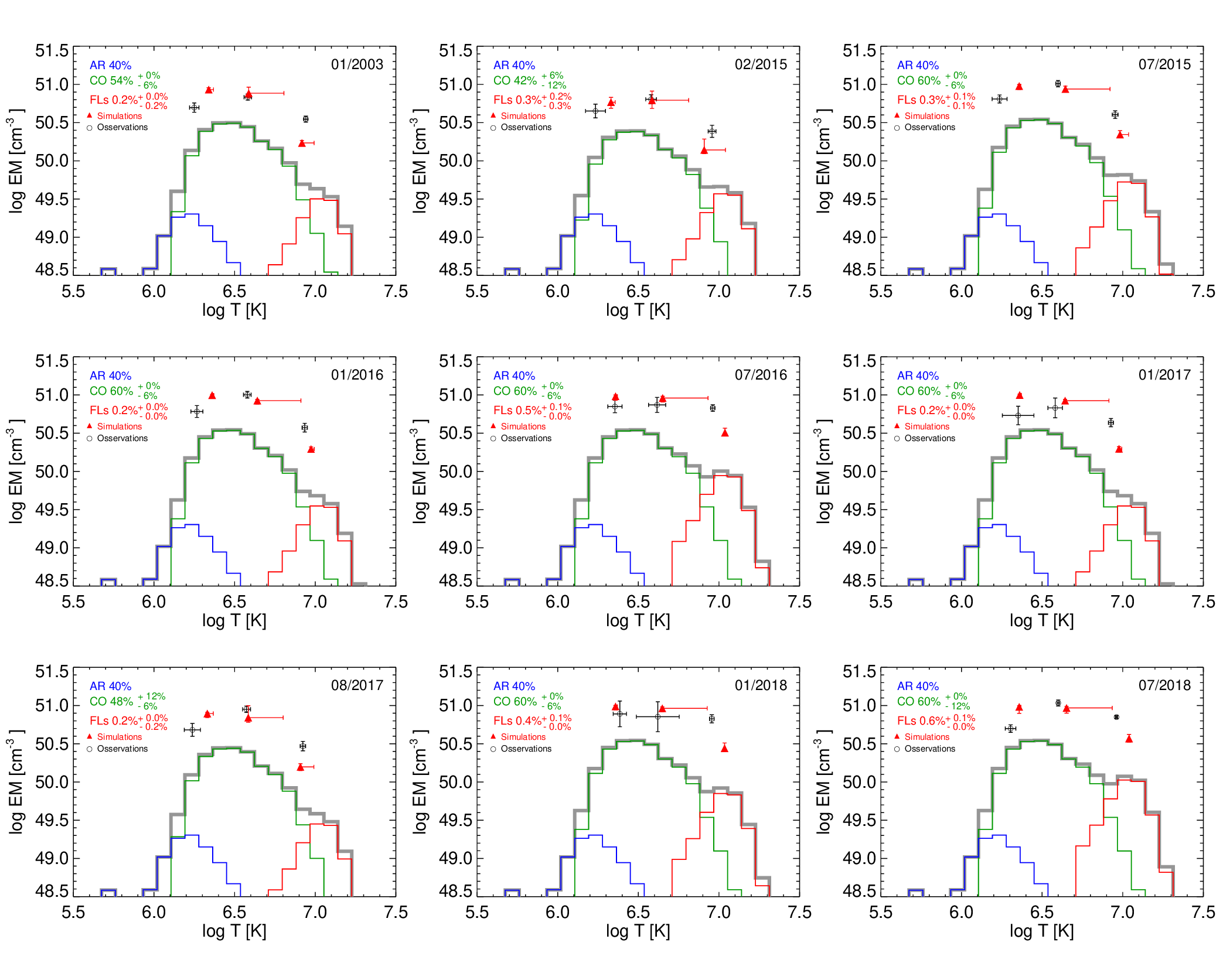}
\caption{Solar coronal EMD (composed of ARs, COs and FLs) that best matches the spectral parameters of $\epsilon \ \rm Eri$. \textit{blue line} - AR contribution; \textit{green line} - CO contribution; \textit{red line} - FL contribution; \textit{grey line} - total EMD of the chosen model. Each observation is represented by a different fraction of surface coverage with AR, CO and FL selected as described in \ref{subsec:std_emd}. \textit{Red triangles} - medians of the 1000 best-fit parameters of the selected synthetic spectra; \textit{black dots} - best-fit parameters of the observations.  }
\label{fig:1stEMD}
\end{figure*}

\begin{figure*}[!htbp]
\includegraphics[width=\hsize, height=8.5cm]{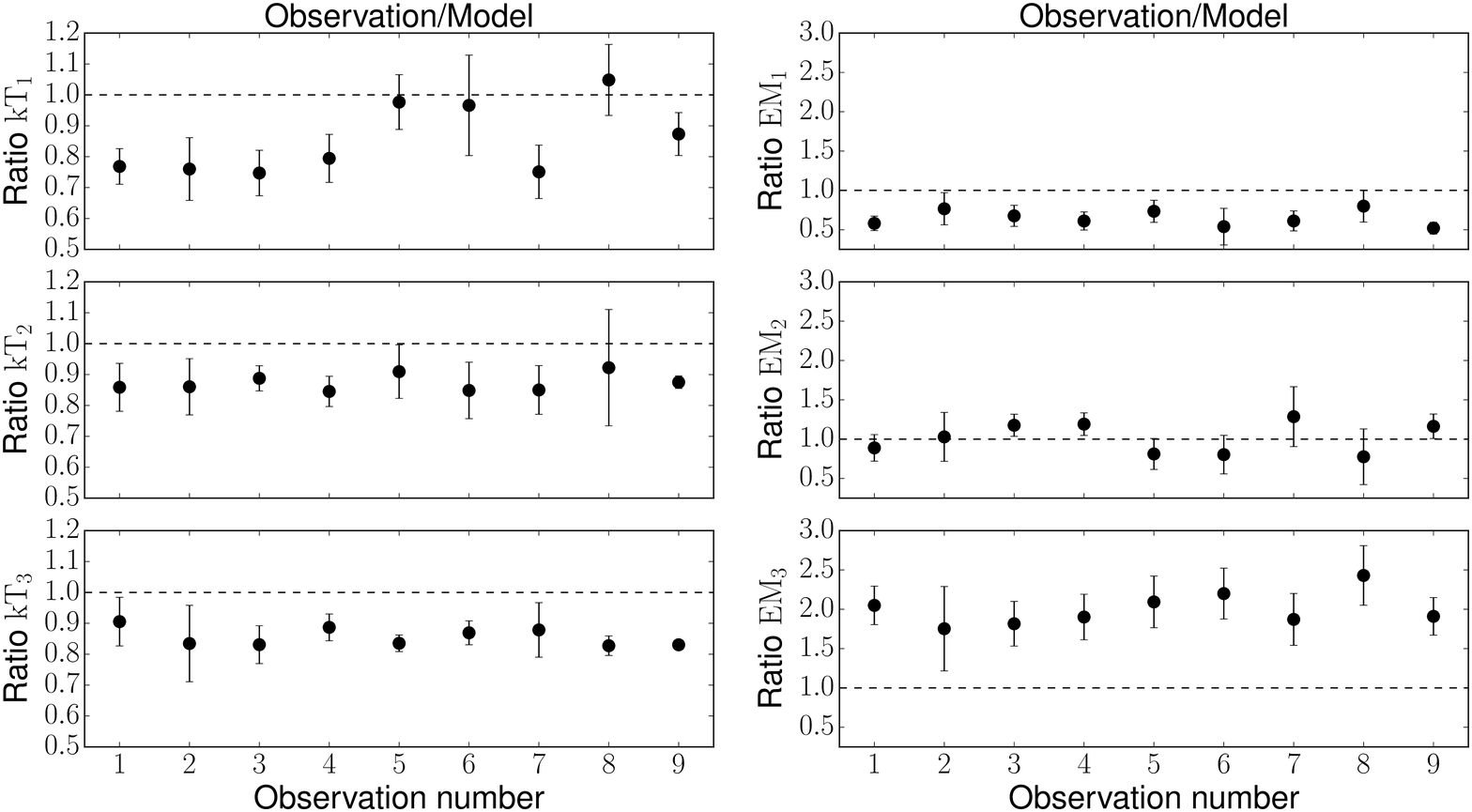}
\caption{Ratio of the best-fit parameters of $\epsilon \ \rm Eri$ over the same parameters selected from the grid with ARs $40 \%$, COs $6-60 \%$ and FLs $0-2 \%$ of COs. The flare EMD is the time averaged distribution of the flares presented by \protect\citet{2001ApJ...557..906R} discussed in \ref{subsec:std_emd}. The error bars are the errors on the ratios found from error propagation taking into account both the uncertainties of observed best-fitting parameters and of the $10 \%$ and $90 \%$ percentiles of the 1000 selected representations.}
\label{fig:ratio}
\end{figure*}
In \ref{fig:ar40_par} we plot the best-fit parameters obtained from the synthetic spectra for the chosen combination of solar structures (case \ref{fig:ar0}c), together with the ones of the best fit to the observed spectra of $\epsilon \ \rm Eri$  (see \ref{tab:spec}).
Analogous to \ref{fig:ar0}, the variation of the colors represents the percent coverage of COs, while the symbol sizes the percentage of FLs. As can be seen from \ref{fig:ar40_par}, an increase of COs on the surface influences the two lower-temperature components of the spectral model, while an increase of the flaring coverage influences mostly the third component, i.e. the hottest one. 

To properly compare the best-fit parameters of the synthetic and the observed spectra, we introduced in the procedure with which we extract the synthetic spectra from the solar-based EMDs a statistical randomization, i.e. Poisson statistics, typical of the satellite's effective area. In this way we simulate the synthetic spectra as if they were actual \textit{XMM-Newton} observations. Moreover, to refine further the comparison, for each grid point we performed a Monte-Carlo simulation: we generated 1000 randomized spectra for each grid point, so that each combination of magnetic structures  is represented by 1000 spectra. We analyzed for each grid point each set of 1000 synthetic spectra in the same way as we had done for the observed spectra of $\epsilon \ \rm Eri$, i.e. we fitted them with a 3-T APEC model with metal abundance fixed at $0.3 \ Z_{\odot}$ (\ref{subsec:specX}). The result of having introduced statistical noise in the simulated spectra is that each combination of magnetic structure is not univocally represented by only one set of values for $kT$ and $EM$, but by 1000 values with a pseudo-random distribution. We demonstrate this in \ref{fig:spread} where, as an example, we show a zoomed plot of the first and the third component of the temperatures. In the plot the black circles are the medians of the 1000 best fit values for each combination of magnetic structures in the range of temperatures displayed in the zoom. For one specific combination (COs $42\%$ and FLs $0.4\%$) we highlight in red its median and, as an example, we overlay in blue the best-fitting temperatures for all 1000 representations of this AR-CO-FL combination. As error bars on the (red) median we adopt the percentile at $10\%$ and $90\%$ of the 1000 values. Evidently, these 1000 values define a significant spread. If this spread is ignored and a one-by-one comparison between observations and synthetic spectra is performed, the result overestimates the truly achievable accuracy in the determination of the best matching model grid point. In other words, the comparison with the synthetic spectra should take into account all the 1000 representations obtained for each grid point.

Next we describe how we performed this match between observed and 
synthetic model parameters. For clarity, in the following we denote
the six spectral parameters ($kT_{1,2,3}$ and EM$_{1,2,3}$) by 
$P_i$, with $i = 1,...,6$. To find the best-fitting combination of 
magnetic structures, we matched the best fit parameters $P_{i}^{obs}$ 
for each observation with each of the corresponding synthetic
parameters $P_{i,j,k}^{syn}$ derived for the 1000 sets (henceforth
labeled with $j$) of all 201 different combinations of solar regions
(grid points; henceforth labeled with $k$), where $j = 1, ... ,
1000$ and $k = 1, ... , 201$. As selection criterion we evaluated
the following six equations:
\begin{equation}
\centering
P_{i}^{obs} - \Delta P_{i}^{obs} \cdot \sigma \le P_{i,j,k}^{syn}
\le P_{i}^{obs} + \Delta P_{i}^{obs} \cdot \sigma
\end{equation}
with a unique parameter $\sigma$ for all six spectral parameters.  
$\sigma$ defines thus the global confidence range of the match between
observed values $P_{i}^{obs}$ and synthetic values $P_{i,j,k}^{syn}$ 
with $i = 1, ..., 6$. We then picked for each of the 1000 sets ($j$)
of combinations of solar regions as best-matching model the one 
among all grid points ($k$) that provides the smallest value for
$\sigma$. The result of this procedure are 1000 best-fitting
representations (for each observation).

The spread of the 1000 representations in terms of spectral parameters (exemplified in \ref{fig:spread}) thus translates into a range of selected best-fitting combinations of AR, CO, and FL coverage. We define as final best-matching combination of ARs, COs and FLs the median of the 1000 values retrieved with our selection procedure and we associate an uncertainty on this result as the $10\%$ and $90\%$ quantile of the 1000 values (red bars in the example of \ref{fig:spread}).
 
The EMDs corresponding to the models selected with this procedure as the best match to the observations are shown in \ref{fig:1stEMD}. The total contribution of all coronal structures is the gray distribution, whereas the blue is the contribution from ARs, the green from COs and the red from FLs. The median area coverage fraction for each of the three types of structures, with the associated errors, is given in the legend of the panels. In addition, the best-fit parameters of the 3-T model fitted to the observed spectra of $\epsilon \ \rm Eri$ with the associated errors from \ref{tab:spec} are plotted (black circles). Finally, in red we show the medians of the best-fit parameters of the 1000 best matching spectra, and the percentiles at $10\%$ and $90\%$ of these representations, denoting the minimum and the maximum of the error bars respectively.

In \ref{fig:ratio} we present a summary of the correspondence between the observed and modeled 3-T best-fitting parameters: we plot the ratio between the best-fit parameters of $\epsilon \ \rm Eri$ and the medians of best-fit parameters of the corresponding selected $EM(T)$.
The ratios of the temperatures result systematically $< 1$, i.e. the temperatures of the selected synthetic spectra are higher than the ones of the observed spectra. 
While the first and the second component of the emission measures give also a ratio $<1$, the $EM_3$ of the selected model is drastically lower than the observed values. Therefore, among the six parameters, the third component of emission measure shows the most drastic discrepancy (see also the discrepancy between the black circles and the red triangles in \ref{fig:1stEMD}).

To summarize, the grid of solar coronal structures with ARs fixed on $40 \%$, COs between $6 \%$ and $60 \%$ and FLs between $0 \%$ and $2 \%$ is able to reproduce the X-ray luminosity and the average coronal temperature of all observed EPIC/pn spectra of $\epsilon \ \rm Eri$ (\ref{fig:ar0}c), but the $EM(T)$ structure does not match very well its spectral shape (\ref{fig:ar40_par}, \ref{fig:1stEMD}, \ref{fig:ratio}).

\subsection{Modified solar coronal structures}
\label{sec:modifiedEMD}
Here we investigate if modifications to the solar coronal EMD can reproduce the observed X-ray spectra of $\epsilon \ \rm Eri$ better.

As shown in \ref{subsec:std_emd}, the most severe discrepancy between synthetic and observed spectra is given by the third component of the $EM$: the synthetic parameters $EM_3$ are systematically lower than the observed ones (\ref{fig:ratio}). The standard EMD previously tested comprises values of $EM_3$ sufficiently high to be compatible with the observed values (see \ref{fig:ar40_par}). However, high values of $EM_3$ in the grid correspond to high values of $kT_3$, that do not agree with the observed $kT_3$, and consequentially are rejected by our selection procedure. Thus, in order to find models that match both the temperatures and the emission measures of the hottest spectral component, we decided to modify the flare $EM(T)$ because FLs are the structures that influence most strongly the hottest component.

\begin{figure}
\centering
\includegraphics[width=\hsize]{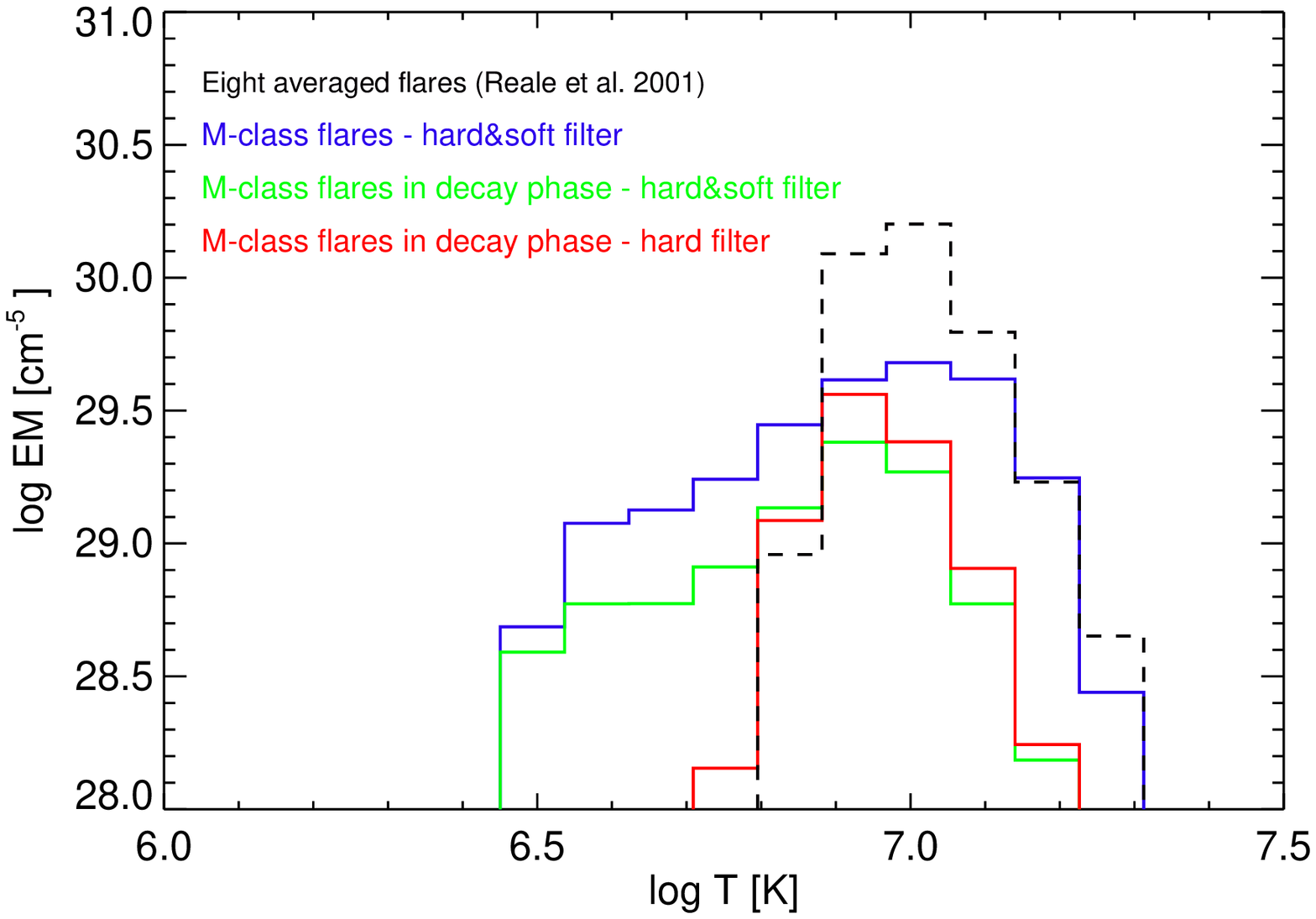}
\caption{Flare $EM(T)$ distributions used to modify the solar coronal EMD, normalized to unit surface. \textit{Black distribution} - time-averaged flares presented in \protect\citet{2001ApJ...557..906R} generated according to \protect\citet{2017A&A...605A..19O}; \textit{blue distribution} - contribution of flares at soft and hard energies;  \textit{green distributions} - contribution of flares in the decay phase; \textit{red distribution} - contribution of flares at the lowest temperatures.}
\label{fig:mix_FL}
\end{figure}

We examined various versions of flare $EM(T)$ as discussed in the following. These $EM(T)$ and the original ones described in \ref{subsec:std_emd} are compared in \ref{fig:mix_FL}.

\subsubsection{Contribution of flares at soft and hard energies}
\label{sec:M_soft_hard}
All the flares presented in \citet{2001ApJ...557..906R} were observed by \textit{Yohkoh} with the two hardest SXT filters, that are sensitive to plasma around and above $10^7$ K. The data published by Reale et al. (2001) comprises two flares of class M that were also observed with softer filters. Soft filter data were published for only one of these two flares, and showed that a contribution from plasma at lower temperatures is important. The soft emission from flares  modifies the flare $EM(T)$ by adding an extended low-temperature tail. We thus replaced the flare $EM(T)$ from \ref{subsec:std_emd} with the average between these two M class flares, including the soft emission. The new flare EMD is the blue distribution in \ref{fig:mix_FL}.

\begin{figure*}[!htbp]
\centering
\includegraphics[width=\hsize]{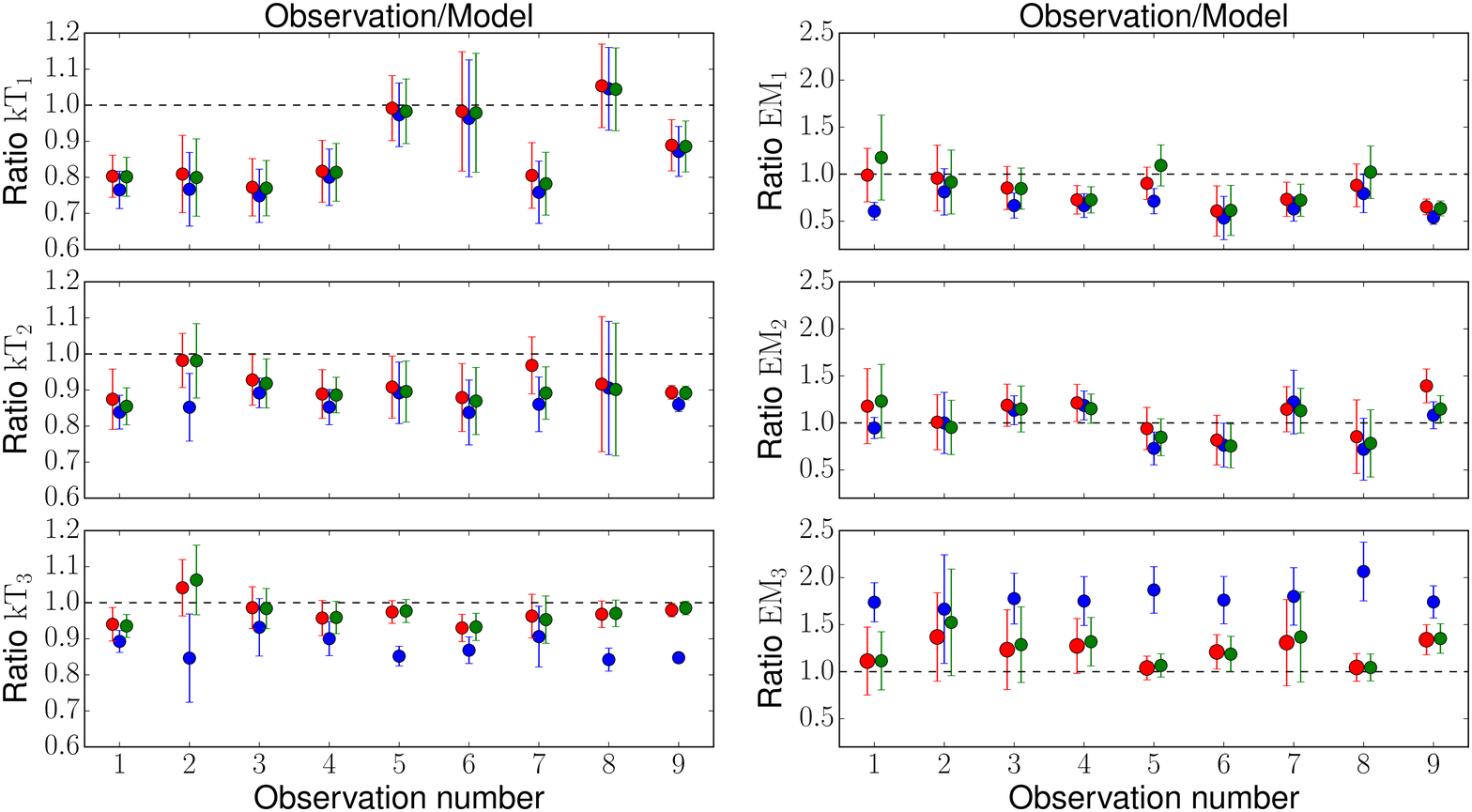}
\caption{Ratio of the best-fit parameters of $\epsilon \ \rm Eri$ over the same parameters selected from the grid with ARs $40 \%$, COs $6-60 \%$. The color code follows \ref{fig:mix_FL}. \textit{Blue circles} - FLs $0-2 \%$ of COs for class M flares at the soft and hard energies; \textit{green circles} - FLs $0-10 \%$ of COs for class M flares at the soft and hard energies in the decay phase; \textit{red circles} - FLs $0-10 \%$ of COs for class M flares at the hard energies in the decay phase. The error bars are the errors on the ratios found from error propagation taking into account both the uncertainties of observed best-fitting parameters and of the $10 \%$ and $90 \%$ percentiles of the 1000 selected representations.}
\label{fig:ratioMsoft}
\end{figure*}

We then built a new grid of EMDs with the same ARs and COs of the previous analysis, i.e. $40 \%$ of ARs and COs varying between $6 \%$ and $60 \%$, and we again set the percentage of FLs between $0 \%$ and $2 \%$ of the percentage of COs. We extracted and fitted the corresponding spectra as in \ref{subsec:std_emd}. \ref{fig:Msoft_hard} shows the EMDs of the corresponding synthetic spectra selected for each observation, analogous to \ref{fig:1stEMD} and with the same meaning of the symbols.
Some of the observations now require a smaller coverage fraction of COs because of the additional soft emission of the flare $EM(T)$. In \ref{fig:ratioMsoft} we plot the ratio between the best-fit parameters of $\epsilon \ \rm Eri$ and the medians of the best-fit parameters of the corresponding selected $EM(T)$ (blue circles). 
The discrepancies between observed and synthetic parameters are somewhat smaller than in the previous case but they show the same pattern (compare blue symbols in \ref{fig:ratioMsoft} with \ref{fig:ratio}).  
In particular for the third component of the emission measure the differences are still large.

\subsubsection{Flares in the decay phase at soft and hard bands}
\label{sec:decay}
We built another flare EM(T) by averaging the two flares of class M observed with the soft filters and the hard filters, but only during their decay phase (green distribution in \ref{fig:mix_FL}). We adopted the same coverage fraction of ARs and COs of the previous analysis, but we set the percentage of FLs to vary between $0 \%$ and $10 \%$ within the percentage of COs. A higher coverage fraction of FLs than the one we adopted previously is required to reproduce the observed $T_{\rm av}$, since this flare EMD has a lower emission measure and a lower temperature compared to the previous EMDs, as can be noticed from \ref{fig:mix_FL}.

We then proceeded on the extraction and fitting of the synthetic spectra and on the selection procedure adopted in \ref{subsec:std_emd}. The EMDs of the corresponding spectra selected for each observation are shown in \ref{fig:Msoft_hard_decay}. These EMDs are composed of significantly lower CO coverage fraction than the previous ones and the FL fraction is up to $\sim 5 \%$. The observed and simulated spectral parameters are now in much better agreement. This is also seen in \ref{fig:ratioMsoft}, where the ratios between the best-fit parameters of $\epsilon \ \rm Eri$ and the medians of the 1000 best-fit parameters of the corresponding selected $EM(T)$ are plotted as green circles. A major improvement is obtained for the hot component, where the ratios for both $kT_3$ and $EM_3$ are now much closer to 1. The parameters representing the other two spectral components also show somewhat better agreement than before. This suggests that flare plasma with lower temperature, such as solar flares during the decay phase, better describes the phenomena in the corona of $\epsilon \ \rm Eri$. 

\subsubsection{Flare in the decay phase at hard band}
\label{sec:dacay_hard}
As final test, we built a new grid of EMDs, using a flare EM(T) of the two class M flares observed only with the hard filters and limited to their decay phase (red distribution in \ref{fig:mix_FL}). The EMDs of the corresponding spectra selected for each observation are shown in \ref{fig:Msoft_decay}. Compared to the flares in decay that included soft emission (\ref{sec:decay}), the FL coverage is smaller and more COs are present. The red circles in \ref{fig:ratioMsoft} are the ratios between the best-fit parameters of $\epsilon \ \rm Eri$ and the medians of the 1000 best-fit parameters of the corresponding selected $EM(T)$. The results are very similar to the previous analysis in \ref{sec:decay}.
The red symbols in \ref{fig:ratioMsoft} are comparable to the green ones, i.e. we can not distinguish which of these two flare distributions better matches the EMD of $\epsilon \ \rm Eri$.

\subsection{Results}
To summarize, among the four different EMDs for solar-like flares, the last two tests involving flares in the decay phase best represent the observed EMD of $\epsilon \ \rm Eri$. In \ref{sec:disc} and \ref{sec:conc} we provide an interpretation of this finding.

\section{Discussion}
\label{sec:disc}
We have detected  for the first time the X-ray cycle of the young solar-like star $\epsilon \ \rm Eri$. Our X-ray monitoring, started in 2015, is longer than the full duration of the known chromospheric Ca\,II H\&K S-index cycle. To validate the newly identified X-ray cycle we have re-analyzed the Ca\,II H\&K S-index variation based on data from 2002 to 2018. The analysis revealed a periodic signal of $2.92 \pm 0.02$ yr, in agreement with past results found on a data set including also historical data \citep{2013ApJ...763L..26M}.  While until 2017 the X-ray variability follows the Ca\,II variability, starting in early 2018 these two activity measures seem to disagree. The chromospheric cycle does not reach the expected maximum, whereas the last X-ray observations show an enhanced X-ray flux. Ongoing continued monitoring of both activity indicators will reveal whether a qualitative change of the cycle is taking place.

\begin{figure}
\includegraphics[width=\hsize]{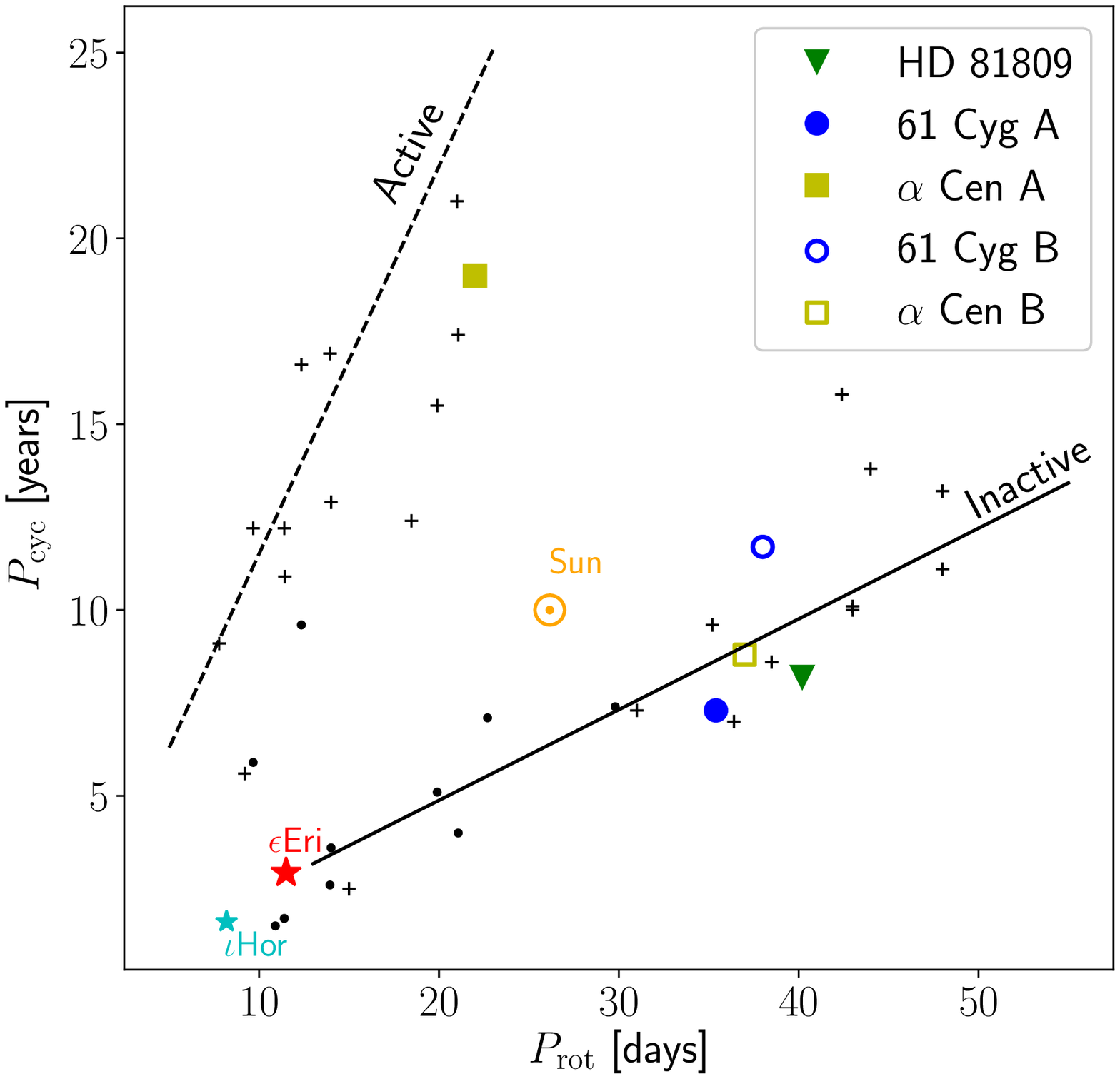}
\caption{Cycle period, $P_{\rm cyc}$, as function of the rotational period, $ P_{\rm rot}$. \textit{Dotted line} - active (A) branch; \textit{solid line} - inactive (I) branch; \textit{black dots and crosses} - original data of \protect\citet{2007ApJ...657..486B}, updated with the new measurements available in \protect\citet{2017ApJ...845...79B}; the black dots are secondary periods of some stars on the A branch; \textit{Colored symbols} - cycles detected in X-rays \protect\citep{2012A&A...543A..84R, 2017A&A...605A..19O, 2013A&A...553L...6S}}
\label{fig:p-p}
\end{figure}

\begin{figure}
\includegraphics[width=\hsize]{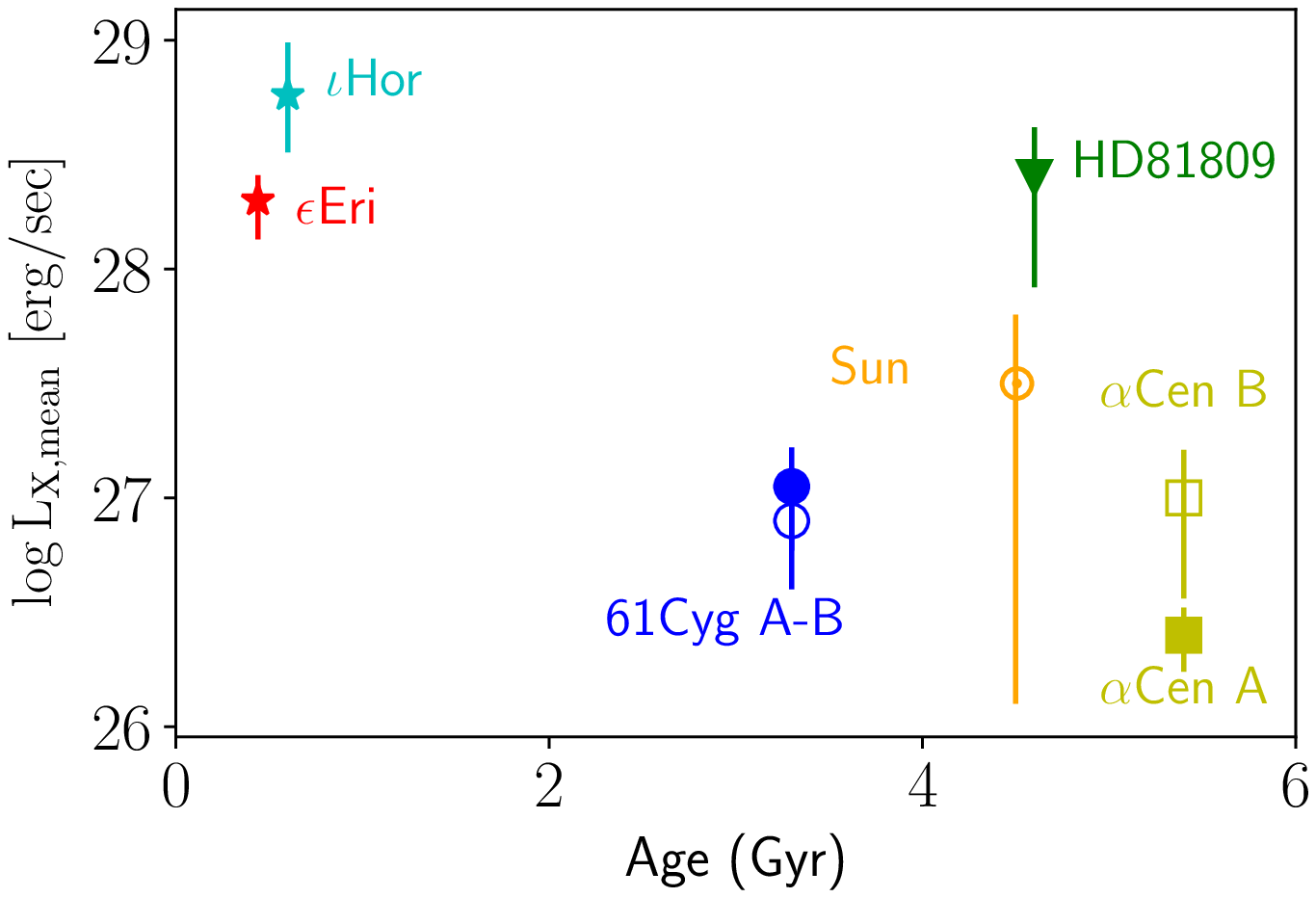}
\caption{The mean X-ray luminosity, $L_{X_{\rm mean}}$, of stars with detected X-ray cycle as function of the age. $L_{X_{\rm mean}}$ is the average of the X-ray available observations of each stars. The vertical bars are the cycle amplitudes.}
\label{fig:age}
\end{figure}
To put $\epsilon \ \rm Eri$ in context to other stars with activity cycles, in particular cycles detected in the X-ray band, we show in \ref{fig:p-p} the relationship between the cycle period $P_{\rm cyc}$ and the rotational period $P_{\rm rot}$, where the updated historical Ca\,II cycles \citep{2007ApJ...657..486B, 2017ApJ...845...79B} are plotted together with the stars with known X-ray activity cycles detected by \textit{XMM-Newton}. The relation between the stellar cycle period, $P_{\rm cyc}$, and the rotation period, $P_{\rm rot}$, provides information on the efficiency of the stellar dynamo. In the $P_{\rm cyc} - P_{\rm rot}$ diagram, a linear relation is found and 
two main branches are distinguished: the so-called active (A) branch, where the dynamo may operate on the surface, and the inactive (I) branch, where  the dynamo may operate in deeper convective zones \citep{1998ApJ...498L..51B,1999ApJ...524..295S,2007ApJ...657..486B,10.1093/mnras/stw2010,  2017ApJ...845...79B, 2018A&A...619A...6O}.
$\epsilon \ \rm Eri$, together with $\iota \rm Hor$, have the shortest X-ray cycles detected so far and are fast rotators, with a rotational period of $11.1$ days \citep{1995ApJ...438..269B} and $8.2$ days \citep{2019A&A...631A..45S} respectively. They are thus placed at the bottom of the inactive branch in the $P_{\rm cyc}-P_{\rm rot}$ diagram. 

In \ref{fig:age} we show the relation between the X-ray luminosity and the stellar age of the stars with confirmed X-ray cycles. The vertical bars in the plot are the amplitude of the cycles, corresponding to the range between the observed maximum and minimum of $L_{\rm X}$.  From our analysis we found that $\epsilon \ \rm Eri$ has an average X-ray luminosity $L_{\rm X}$ of $2.0 \times 10^{28} \ \rm erg/s$, with an amplitude of $1.3 \times 10^{28} \ \rm erg/s$. The values of the other X-ray luminosities and their amplitudes are taken from \citet{2012A&A...543A..84R} for the $\alpha$ Cen and 61 Cyg systems, from \citet{2017A&A...605A..19O} for HD~81809 and from \citet{2013A&A...553L...6S} for $\iota$ Hor. Consistent with the decrease of X-ray luminosity with stellar age, the two youngest stars, $\epsilon \ \rm Eri$ and $\iota$ Hor, have the highest X-ray luminosity of all stars with known X-ray cycles. We notice also that the amplitudes of their cycles have the lowest values. 

In this work, we provide a physical explanation for these findings thanks to a new tecnique to interpret the stellar X-ray spectra. This method allows us to examine the evolution of the corona of $\epsilon \ \rm Eri$ during its X-ray cycle in terms of the EMDs of solar magnetic structures, i.e. active regions, cores of active regions and flares (ARs, COs and FLs). The same method had been applied before only to one star, HD 81809, by \citet{2008A&A...490.1121F} and \citet{2017A&A...605A..19O}. 

Compared to HD 81809, the X-ray spectra of $\epsilon \ \rm Eri$ are of much higher quality and, therefore, they provide more information on the coronal temperature structure. This requires a more sophisticated analysis of the effects of the statistical noise on the accuracy with which the spectral parameters can be constrained. Thus, unlike to the literature studies of HD~81809, we performed a Monte-Carlo simulation for each combination of magnetic structures. As we showed for an example in \ref{fig:spread},
the spectral parameters retrieved
from fitting the synthetic spectra including random statistical noise show fluctuations that are
larger than the typical separation of individual grid points
representing specific combinations of ARs, COs and FLs. This leads to a degeneracy in the mapping of spectral parameters ($kT$ and $EM$) to the magnetic region coverage fraction. Therefore,
a one-to-one comparison between the observed X-ray spectra of  $\epsilon \ \rm Eri$ 
and the synthetic spectra, such as done in previous works on HD~81809,
overestimates the achievable accuracy. We
have instead here determined the range of uncertainty associated with each
best-fitting model (represented by a specific percentage of ARs, COs and FLs) from a statistical assessment of the above-mentioned fluctuations, as demonstrated in \ref{fig:spread}.

As a result of our detailed analysis, we found that the emission measure distributions of $\epsilon \ \rm Eri$ match the solar templates only when we assume flares that are not representative of the solar average flare distribution. The best match with the observed X-ray spectra of $\epsilon \ \rm Eri$ is found for flare $EM(T)$ representing solar flares during their decay. In this phase flares are cooling and, therefore, their EMD results shifted to lower temperatures with respect to the time-averaged solar flare EMD. A possible interpretation of this finding is that the actual flare distribution on $\epsilon \ \rm Eri$ at any given time is dominated by flares in their late stage of evolution. Such a scenario implies that the typical duration of X-ray flares on $\epsilon \ \rm Eri$ is longer than the duration of their solar counterparts. This could be due to the lower metallicity, compared to the Sun, that makes radiative losses less efficient.

\begin{figure}
\centering
\includegraphics[width=\hsize]{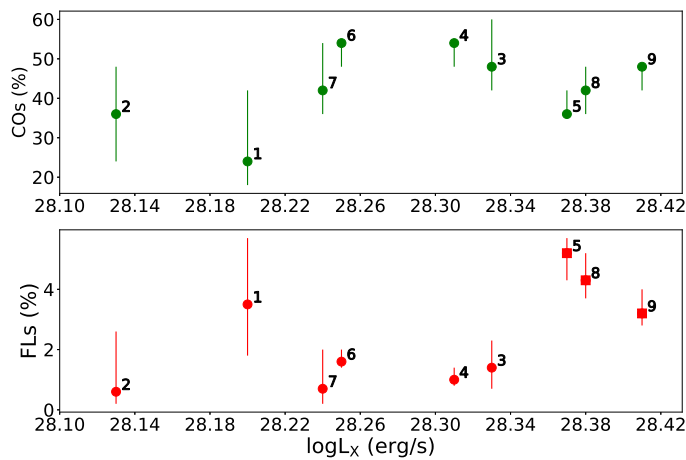}
\caption{Coverage fraction of COs and FLs as function of the X-ray luminosity of $\epsilon \ \rm Eri$. The top panel refers to the variation of percentage of COs that cover the surface of $\epsilon \ \rm Eri$ throughout its cycle. The bottom panel shows the variation in percentage of the flaring component. The squares correspond to the observations with a flaring event in the short-term lightcurve. The error bars are the percentile at $10\%$ and $90\%$ of the 1000 best-matching representations selected for each observation.}
\label{fig:perc}
\end{figure}

In \ref{fig:perc} we show the coverage fraction of COs and FLs throughout the cycle of $\epsilon \ \rm Eri$ as function of its X-ray luminosity. The percentages of COs and FLs are the ones from our best-matching trial EMD, i.e. the one from \ref{sec:decay}, where the flares are in the decay phase and both hard and soft filters were used in the \textit{Yohkoh} observations (flare EMD in green in \ref{fig:mix_FL})~\footnote{While the test where the flare EMD is restricted to the use of the hard filter (red distribution in \ref{fig:mix_FL} (see \ref{sec:dacay_hard}) provides similar results, the exclusion of the soft flare component is artificial and we do not adopt this as the final EMD.}. 
During the minima of the cycle (January 2003 and February 2015) the coverage of cores of active regions is $24 \%$ and $36 \%$ respectively, while in the other phases of the cycle it increases up to $54 \%$. Outside the cycle minimum the cores are thus the dominating magnetic structures in terms of spatial coverage of the corona of  $\epsilon \ \rm Eri$. The maxima of the cycle require an increased percentage of flares: $\sim 3 \%$ to $\sim 5 \%$, compared to $< 2\%$ for states of lower $L_{\rm X}$. This is in agreement with the short-term lightcurves that show flare-like variability predominately during the cycle maximum (see \ref{sec:lc} and figures in Appendix~A). Thus, the flares are superposed on the cyclic variability in the X-ray waveband and therefore they have to be disentangled from the cycle variations.

In the solar cycle the flaring component observed on the Sun influences weakly the total solar corona having a marginal contribution throughout the cycle, as well as the cores of active regions that result absent in the cycle minimum and feeble in the cycle maximum \citep{2001ApJ...560..499O}. Instead, for the cycle of $\epsilon \ \rm Eri$ we have shown that during the minimum $44-76\%$ of the total surface results to be covered by magnetic structures ($40 \%$ ARs and $24-36 \%$ COs), going up to $ 88\%$ in the maximum ($40 \%$ ARs and $48 \%$ COs).

\begin{figure}
\centering
\includegraphics[width=\hsize]{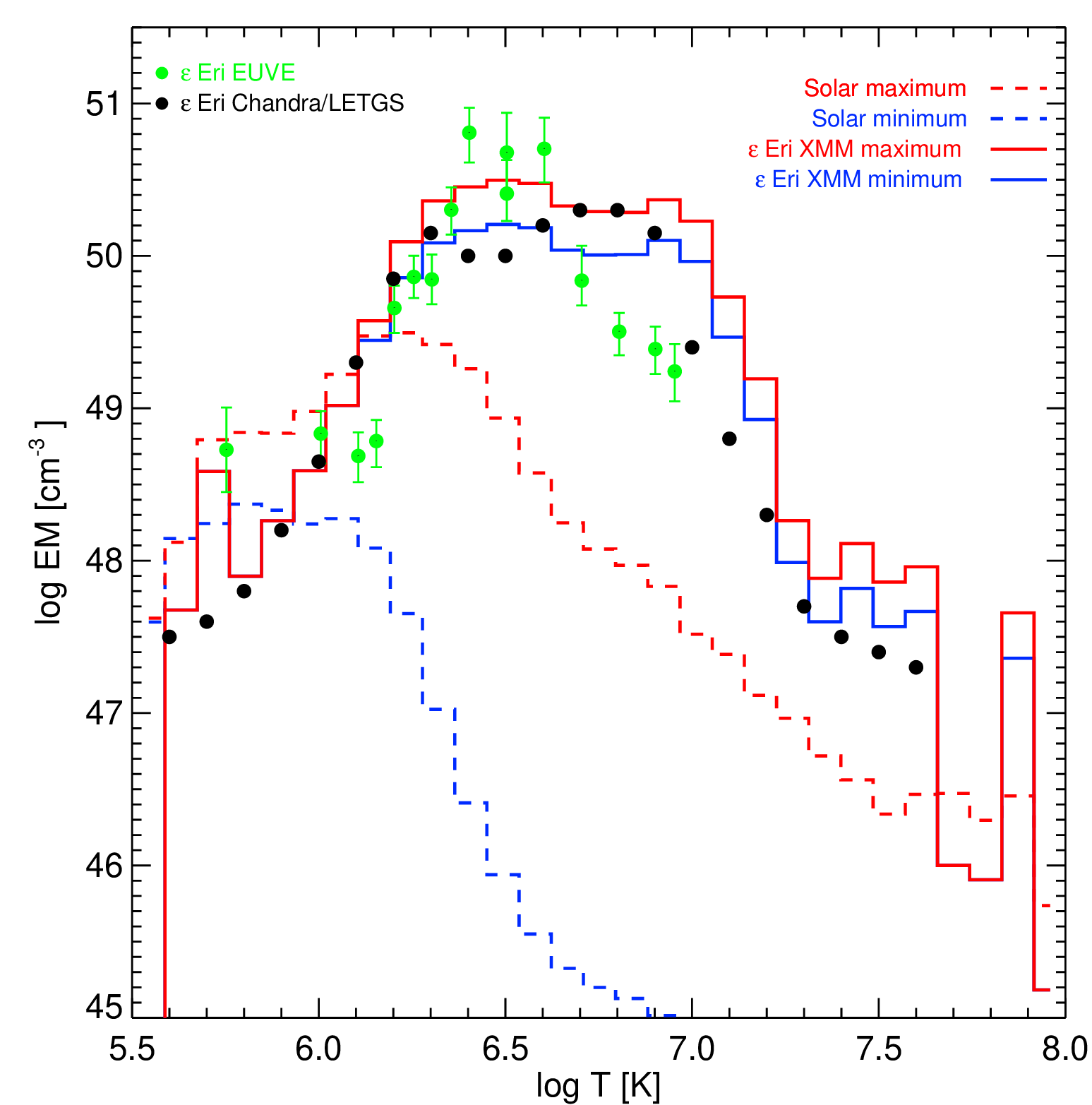}
\caption{EMDs of the Sun and $\epsilon \ \rm Eri$. \textit{Dashed lines}: solar EMDs during the minimum in April 1996 (blue) and the maximum in December 1991 (red) of the solar cycle; \textit{solid lines}: EPIC/pn EMDs of $\epsilon \ \rm Eri$ during the minimum in February 2015 (blue) and maximum in July 2018 (red) of its X-ray cycle; \textit{green dots}: EMD of $\epsilon \ \rm Eri$ from the EUVE spectra of 1993 \protect\citep{2000ApJ...545.1074D}; \textit{blacks dots}: EMD of $\epsilon \ \rm Eri$ from \textit{Chandra/LETGS} spectrum taken in March 2001 \protect\citep{2004A&A...416..281S}.}
\label{fig:emd_confr}
\end{figure}

In the past \citet{2000ApJ...545.1074D} directly compared the emission measure distribution of the solar active regions extracted from \textit{Yohkoh} images with the EMD found for $\epsilon \ \rm Eri$ with \textit{EUVE} spectra obtained in 1993. \citet{2003ApJS..145..147S, 2004A&A...416..281S} also analyzed EMDs of $\epsilon$~Eri using \textit{EUVE} and the X-ray satellite \textit{Chandra}, deriving a similar shape to that of \citet{2000ApJ...545.1074D}
 in the case of the \textit{EUVE} data. Both these studies are based on an analysis of the high resolution spectra of $\epsilon \ \rm Eri$. In \ref{fig:emd_confr} we compare these EMDs (green and black dots) to the emission measure distributions of the Sun during the minimum and the maximum of the solar cycle (April 1996 and December 1991; dashed distributions) and the EMDs of $\epsilon \ \rm Eri$ during the minimum and the maximum of the X-ray cycle derived by us with an entirely different method (February 2015 and August 2018; solid distributions).
The EMDs found with our method are in good agreement with the X-ray EMD derived by \citet{2004A&A...416..281S}, strengthening the validity of our analysis that, even if we used low-resolution spectra, is able to reproduce past results based on high-resolution spectra. At low temperatures ($T \sim 10^6$ K), we reproduce well the emission measure of $\epsilon \ \rm Eri$ presented by \citet{2000ApJ...545.1074D} and its rise towards higher temperatures that reach a first peak at $T \sim 10^{6.5} \ \rm K$. At high temperatures the X-ray emission measures show a different maximum than the EUVE data. The peaks at $\sim 10^7$ K in the X-ray emission measures found in our analysis are the contribution of the flaring events. The absence of this peak in the EMD obtained with \textit{EUVE} shows that these data have no
sensitivity for the flaring component. Moreover, at the time of \textit{EUVE} observations in 1993 $\epsilon \ \rm Eri$ did not show the $3$-yr chromospheric cycle, i.e. the EMD could have been different, and thus it is reasonable to think that the current corona of $\epsilon$~Eri has more flaring events than it had 26 years ago. 
It can be seen from \ref{fig:emd_confr} that the EMDs of $\epsilon \ \rm Eri$ are very different from those of the Sun throughout its activity cycle. As noticed by \citet{2000ApJ...545.1074D}, these difference may lay on the fact that active stars may be covered with more active regions than the Sun. With our method we confirmed that the high X-ray emission from $\epsilon \ \rm Eri$ can, indeed, be explained by a high surface coverage of magnetic structures.   

\section{Conclusions}
\label{sec:conc}
We have analyzed in detail the long-term X-ray variability of $\epsilon \ \rm Eri$ and we showed evidence for an X-ray cycle for the first time.
With an age of $440$ Myr, $\epsilon \ \rm Eri$, together with $\iota$ Hor, are the youngest stars with detected X-ray activity cycles. 

We have applied a new method in which we describe the X-ray spectra of $\epsilon \ \rm Eri$ and the evolution of the X-ray cycle in terms of solar emission measure distributions considering active regions, cores of active regions and flares in the corona. We found that during the minimum of the cycle, $\sim 60 \%$ of the surface of $\epsilon \ \rm Eri$ is covered by these structures, going up to $\sim 90 \%$ throughout the X-ray cycle. We have thus found direct evidence that the high X-ray luminosity of $\epsilon \ \rm Eri$, and likely of young fast rotating stars in general, are the result of high magnetic filling factor in the corona. This result is also bolstered by the small amplitudes of the X-ray cycle of $\epsilon \ \rm Eri$, and the similarly young star $\iota$ Hor, which - in the vein of the above arguments - can be explained as a lack of additional space for enhancing the covering fraction throughout the cycle. 
Therefore, this suggests that an age of $\sim 400$ Myr is the youngest age for coronal X-ray cycles to set in as in even younger stars the basal surface coverage with active structures is likely to be even higher. 

We found that the corona of $\epsilon \ \rm Eri$ can be described in terms of solar magnetic structures only if the standard solar flare EMD is replaced by an EMD representing exclusively the decay phase of flares. We conclude that the flaring events that take place on the surface of $\epsilon \ \rm Eri$  last longer than typical solar flares and we ascribe this to the low metallicity of $\epsilon \ \rm Eri$ which slows down the radiative cooling in the corona.

$\epsilon \ \rm Eri$ is now entering in an interesting state in which X-ray and Ca\,II cycle are not as well correlated as in the years before: while the long-term X-ray lightcurve seems to indicate an anticipation of the latest cycle maximum, the Ca\,II $S_{\rm MWO}$-index variability has lately strongly decreased. Taken together with the historical Ca\,II observations (not considered in this work) where different cycle periods were found, it is clear that activity cycles on $\epsilon \ \rm Eri$ are not a stable phenomenon. Our continued Ca\,II and X-ray monitoring of this star will shed more light on this issue.

Our study shows that X-ray cycles can be present on young stars albeit with different characteristics than the solar cycle. To better validate this evidence, continuing to monitor the X-ray activity of such targets is particularly important.

\begin{acknowledgements} 
MC and LD acknowledge financial support by the Bundesministerium f\"ur Wirtschaft und Energie through the Deutsches Zentrum f\"ur Luft- und Raumfahrt e.V. (DLR) under the grants FKZ 50 OR 1708 and FKZ 50 OG 1602.
The research leading to these results has received also funding from the European Union's Horizon 2020 Program under the AHEAD project (grant agreement 654215). T.S.M. acknowledges the support of grant AST-1812634 from the U.S. National Science Foundation.
UW acknowledges funding by DLR, project FKZ 50 OR 1701, using data obtained with the TIGRE telescope, located at La Luz observatory, Mexico. TIGRE is a collaboration of the Hamburger Sternwarte, the Universities of Hamburg, Guanajuato and Li\`ege. JSF acknowledges
support from the Spanish MICINN through grant AYA2016-79425-C3-2-P.
\end{acknowledgements}

\bibliographystyle{aa} 
\bibliography{eeri_20_v2} 

\begin{appendix}
\label{appendix:shot_lc}

\section{EPIC/pn lightcurve}
The EPIC/pn lightcurves are shown. The time bin size chosen is set on 300 s. The solid red lines overplotted on each lightcurve represent the segmentation found with the software \textit{R}. 
\begin{figure}[!ht]
\centering
\includegraphics[width=0.8\columnwidth]{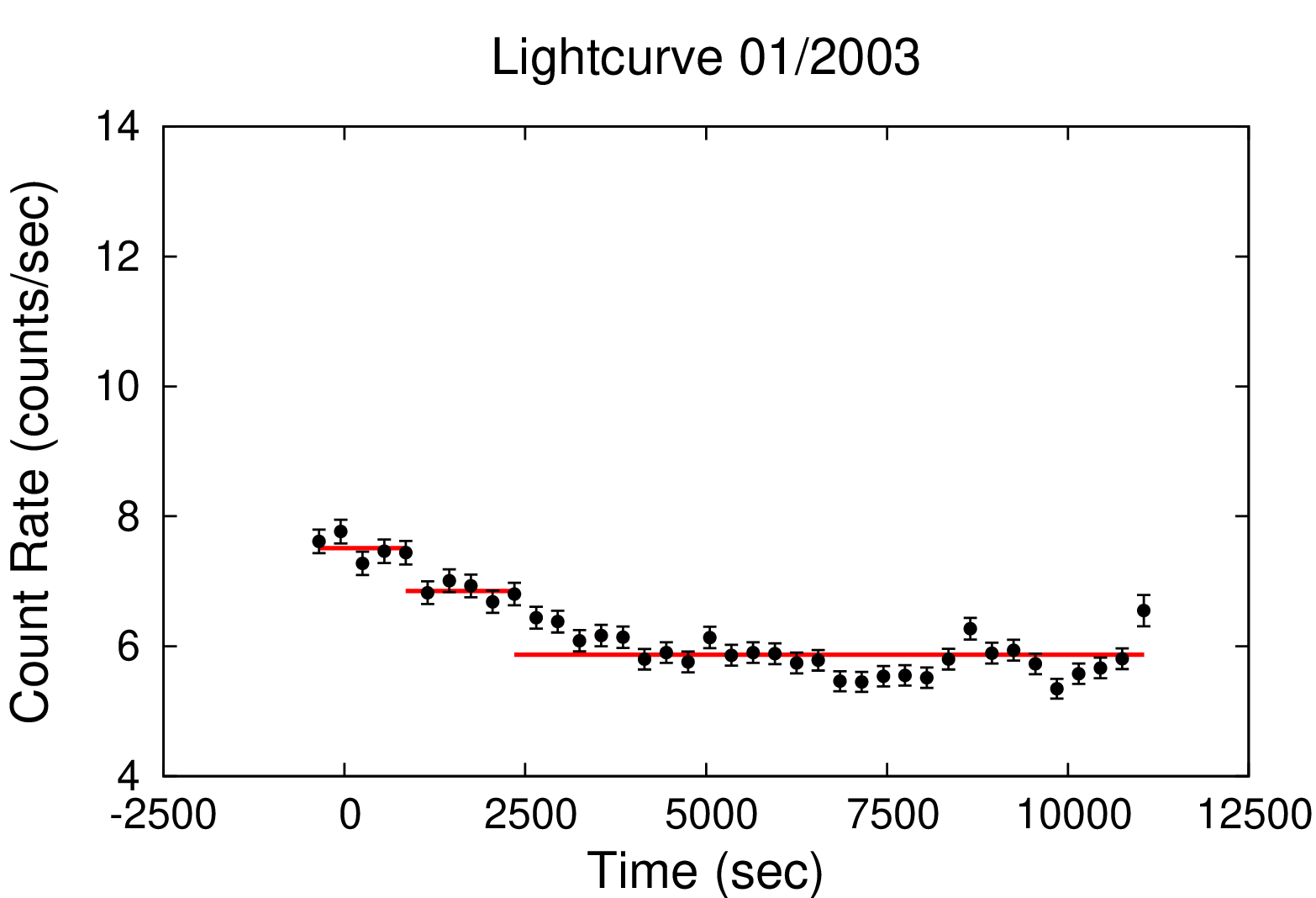}
\end{figure}
\begin{figure}[!ht]
\centering
\includegraphics[width=0.8\columnwidth]{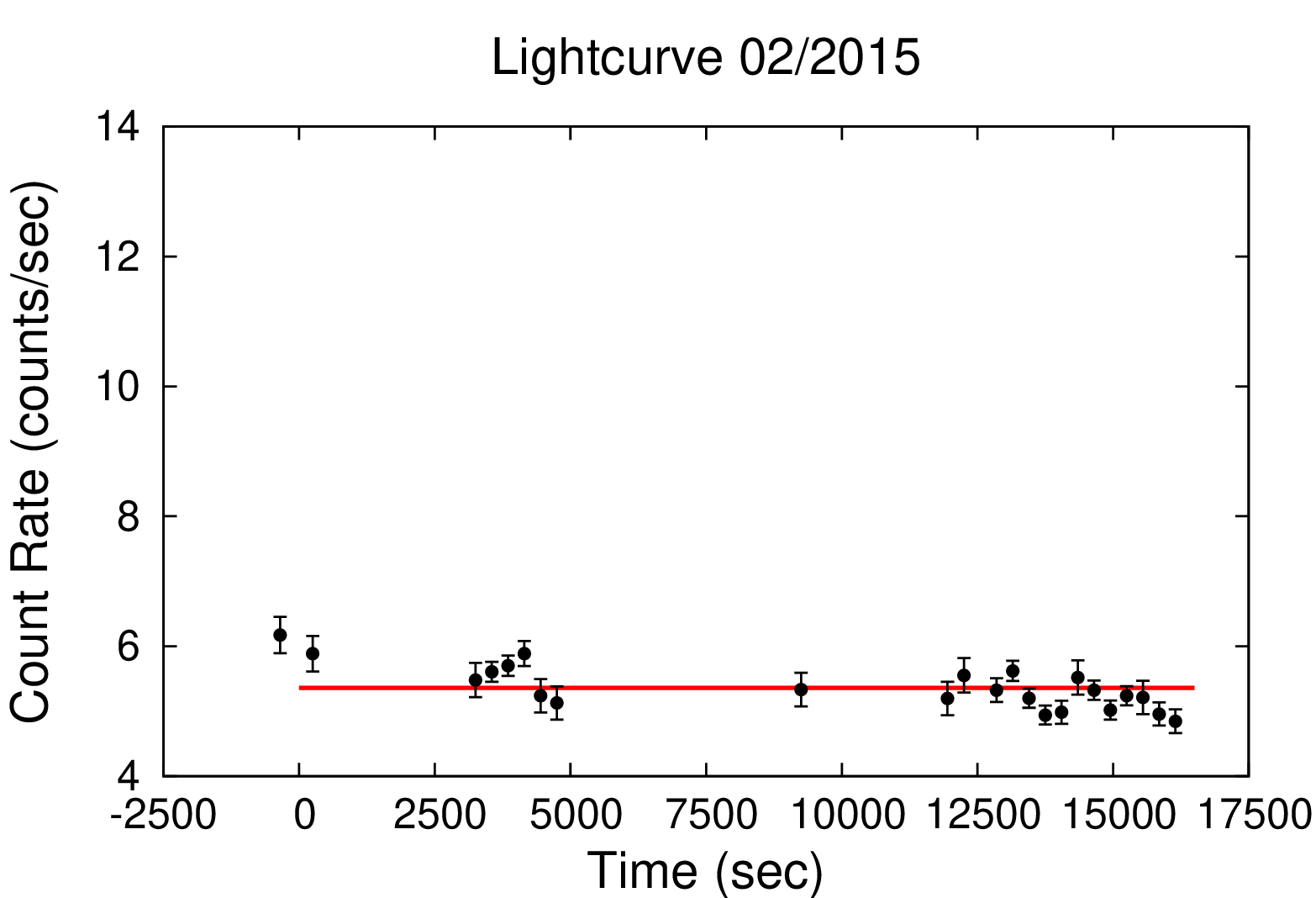}
\end{figure}
\begin{figure}[!ht]
\centering
\includegraphics[width=0.8\columnwidth]{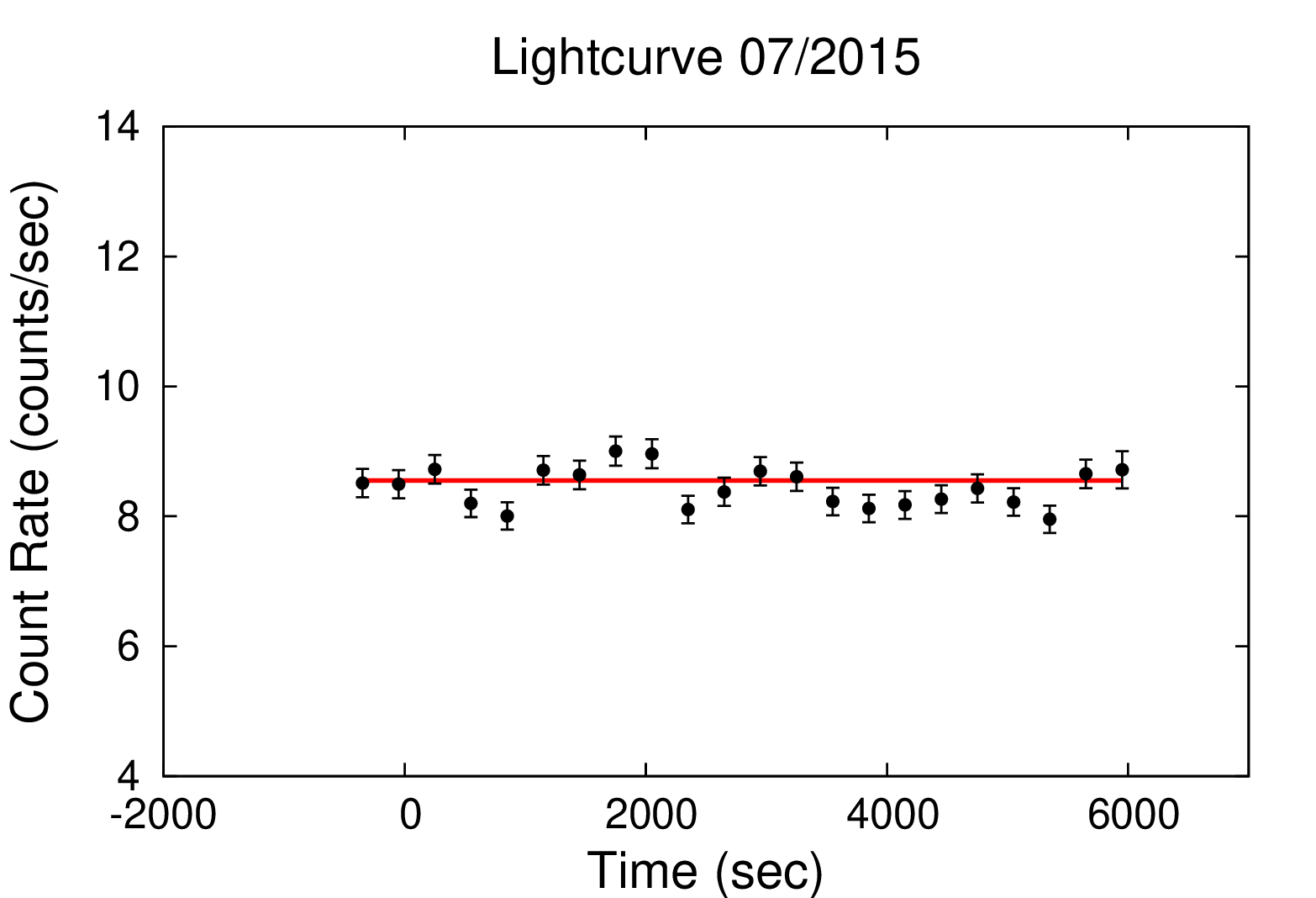}
\end{figure}
\begin{figure}[!ht]
\centering
\includegraphics[width=0.8\columnwidth]{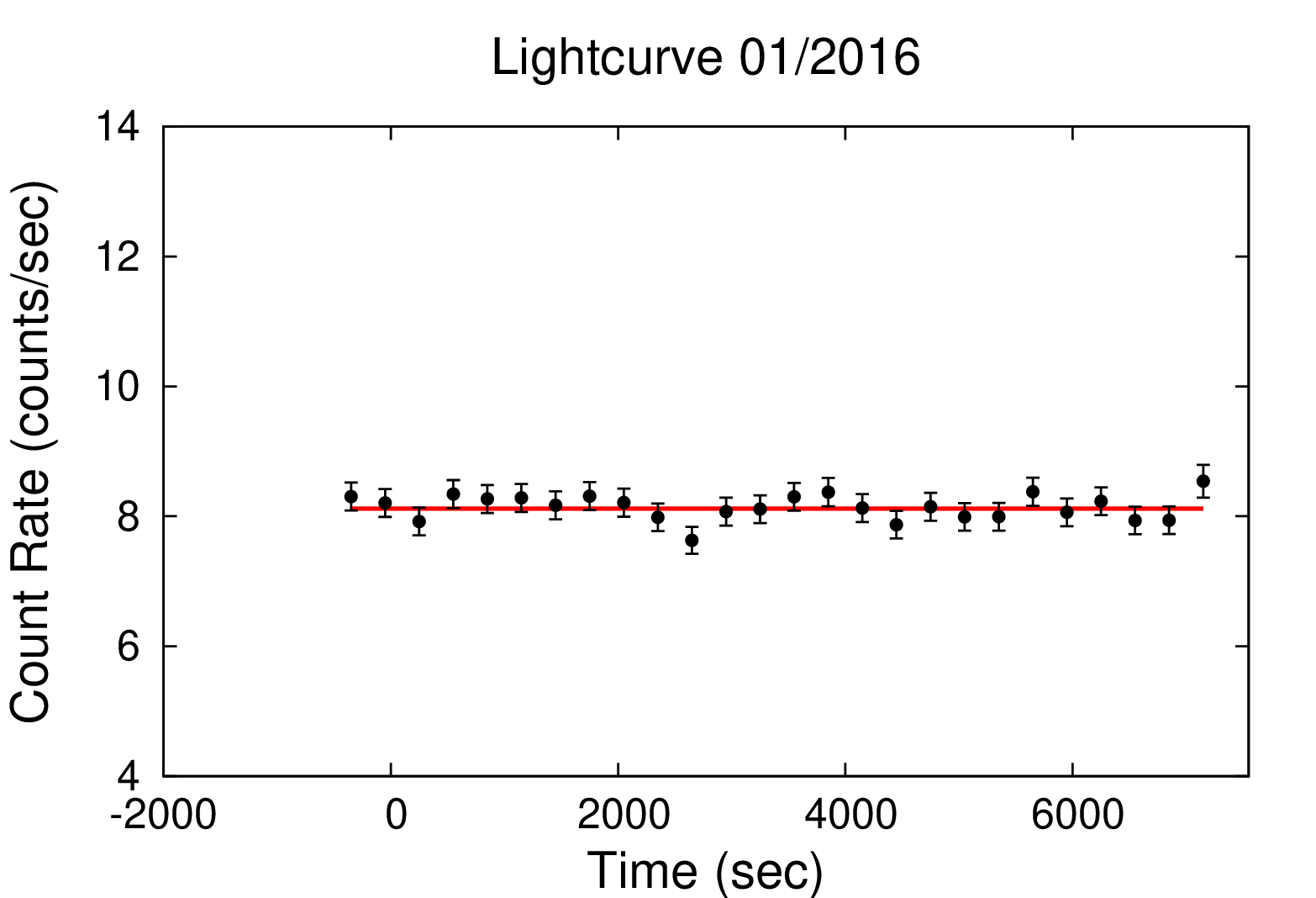}
\end{figure}
\begin{figure}[!ht]
\centering
\includegraphics[width=0.8\columnwidth]{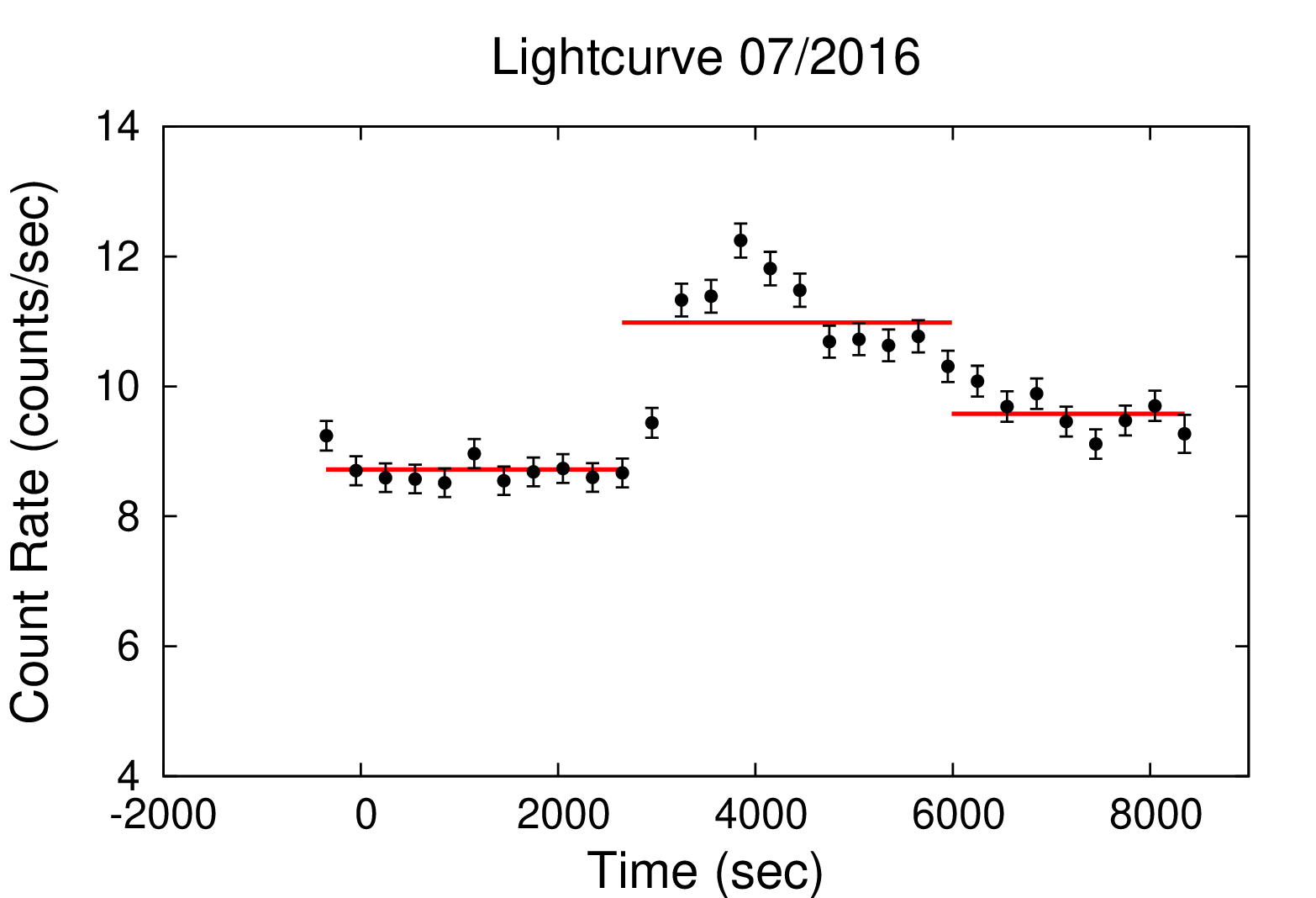}
\end{figure}
\begin{figure}[!ht]
\centering
\includegraphics[width=0.8\columnwidth]{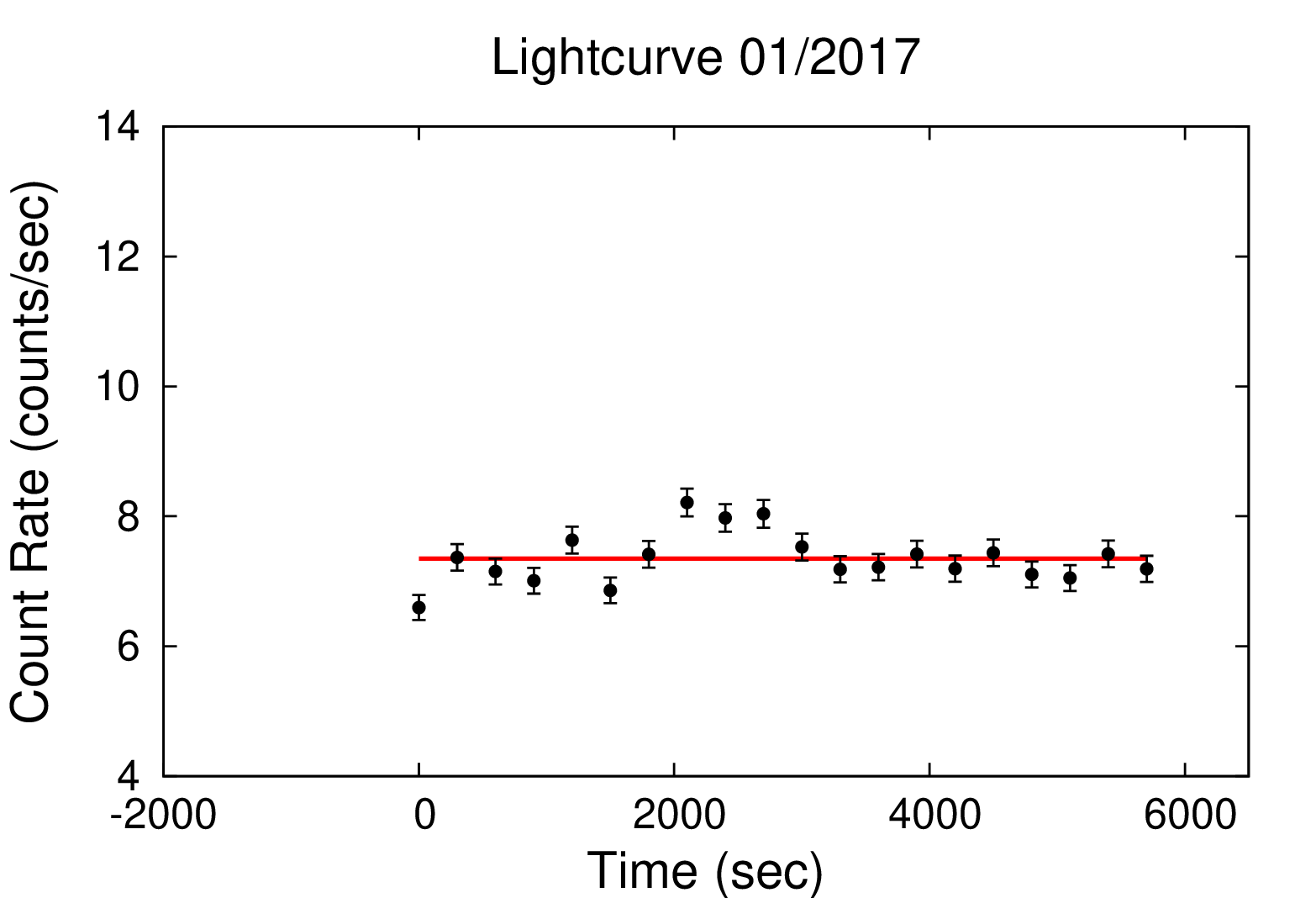}
\end{figure}
\begin{figure}[!ht]
\centering
\includegraphics[width=0.8\columnwidth]{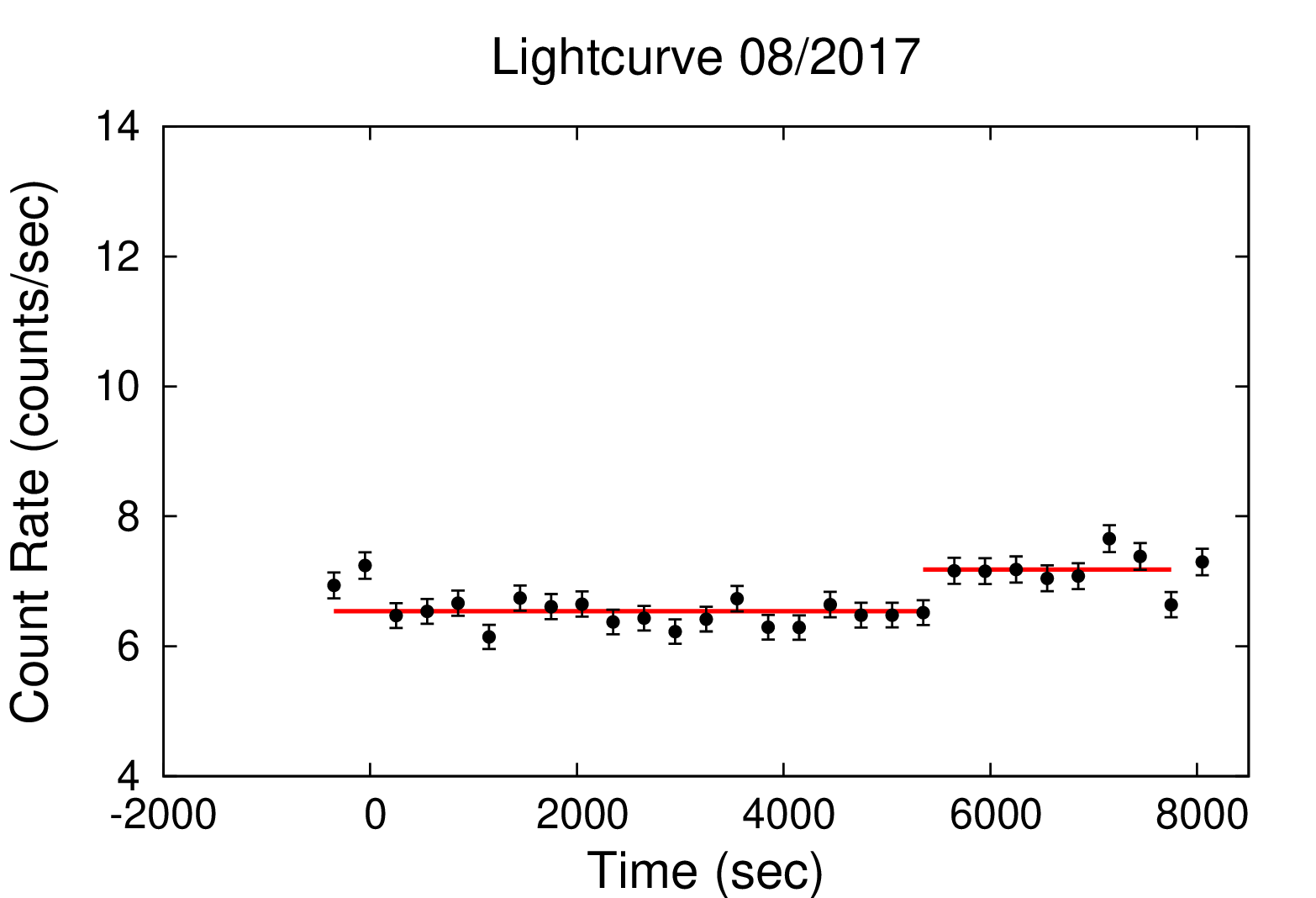} 
\end{figure}
\begin{figure}[!ht]
\centering
\includegraphics[width=0.8\columnwidth]{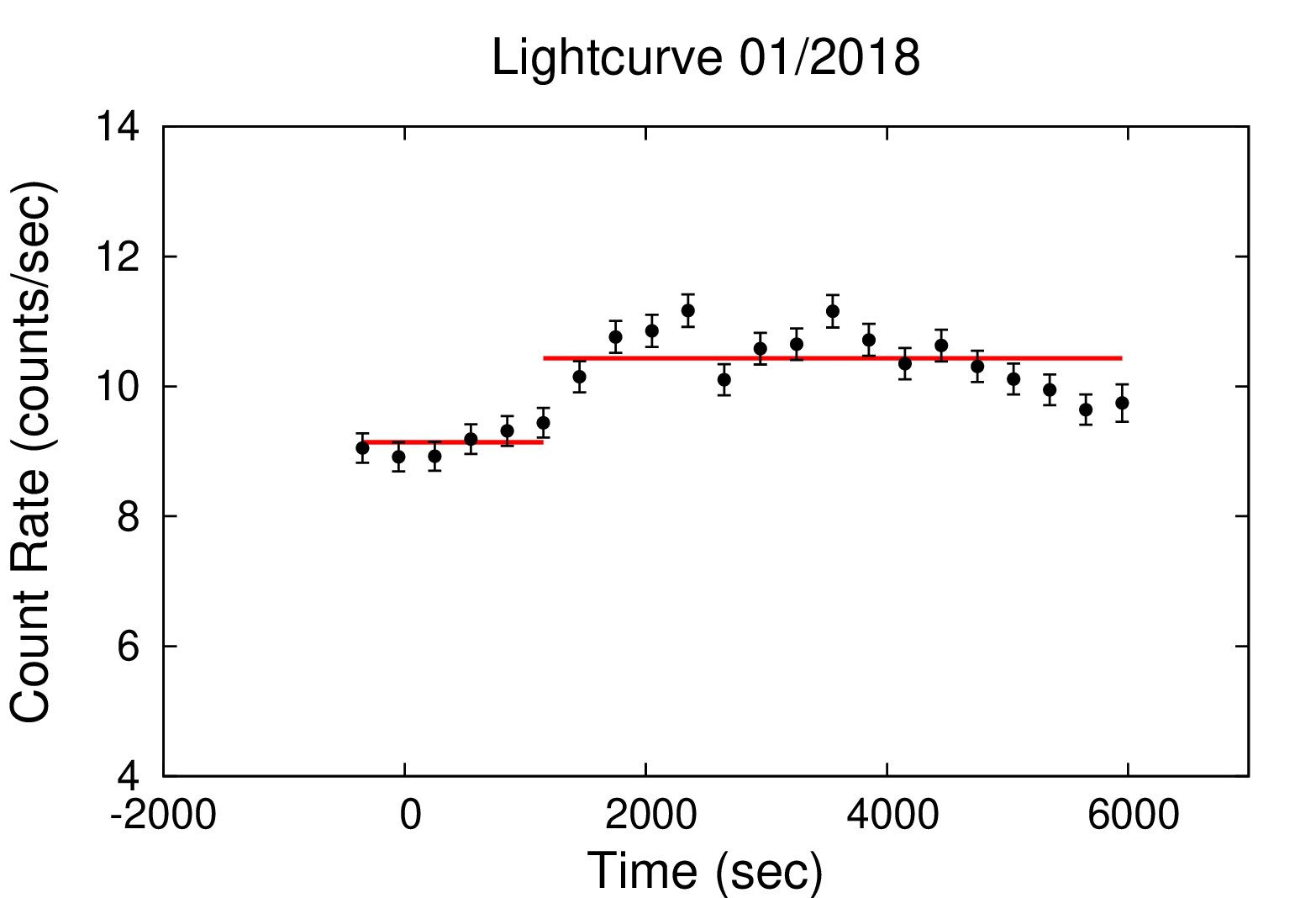}
\end{figure}
\begin{figure}[!ht]
\centering
\includegraphics[width=0.8\columnwidth]{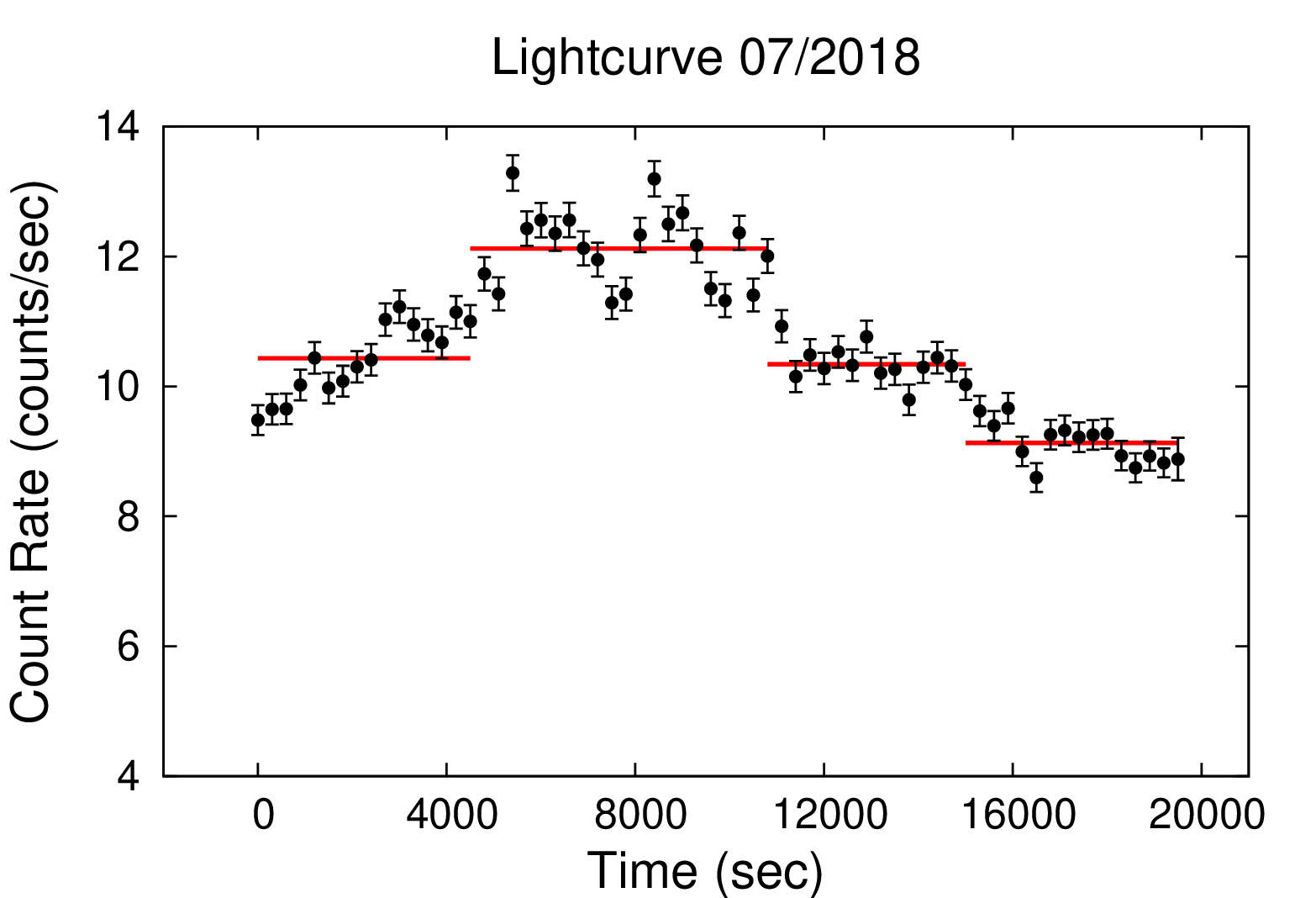}
\end{figure}

\clearpage

\section{Modified solar coronal EMD}
The EMDs selected according to the criterion given in \ref{subsec:std_emd} are shown for the three modified flare EM(T) considered in this work and discussed in \ref{sec:modifiedEMD}.

\begin{figure*}
\centering
\includegraphics[width=\hsize]{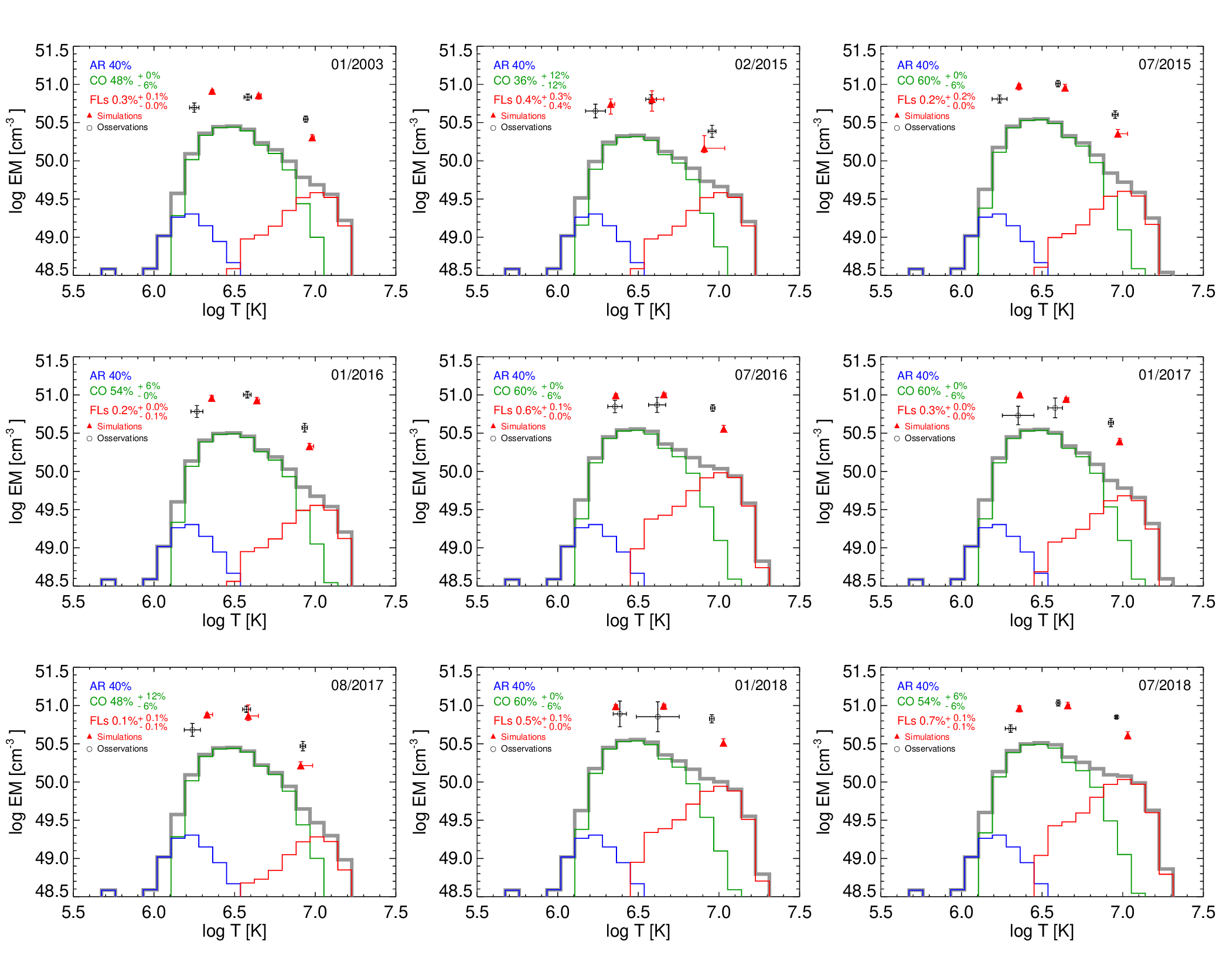}
\caption{Selected best-fitting EMD for class M flares at the soft and hard energies. The coding of the plots is the same as in \ref{fig:1stEMD}.}
\label{fig:Msoft_hard}
\end{figure*}

\begin{figure*}
\centering
\includegraphics[width=\hsize]{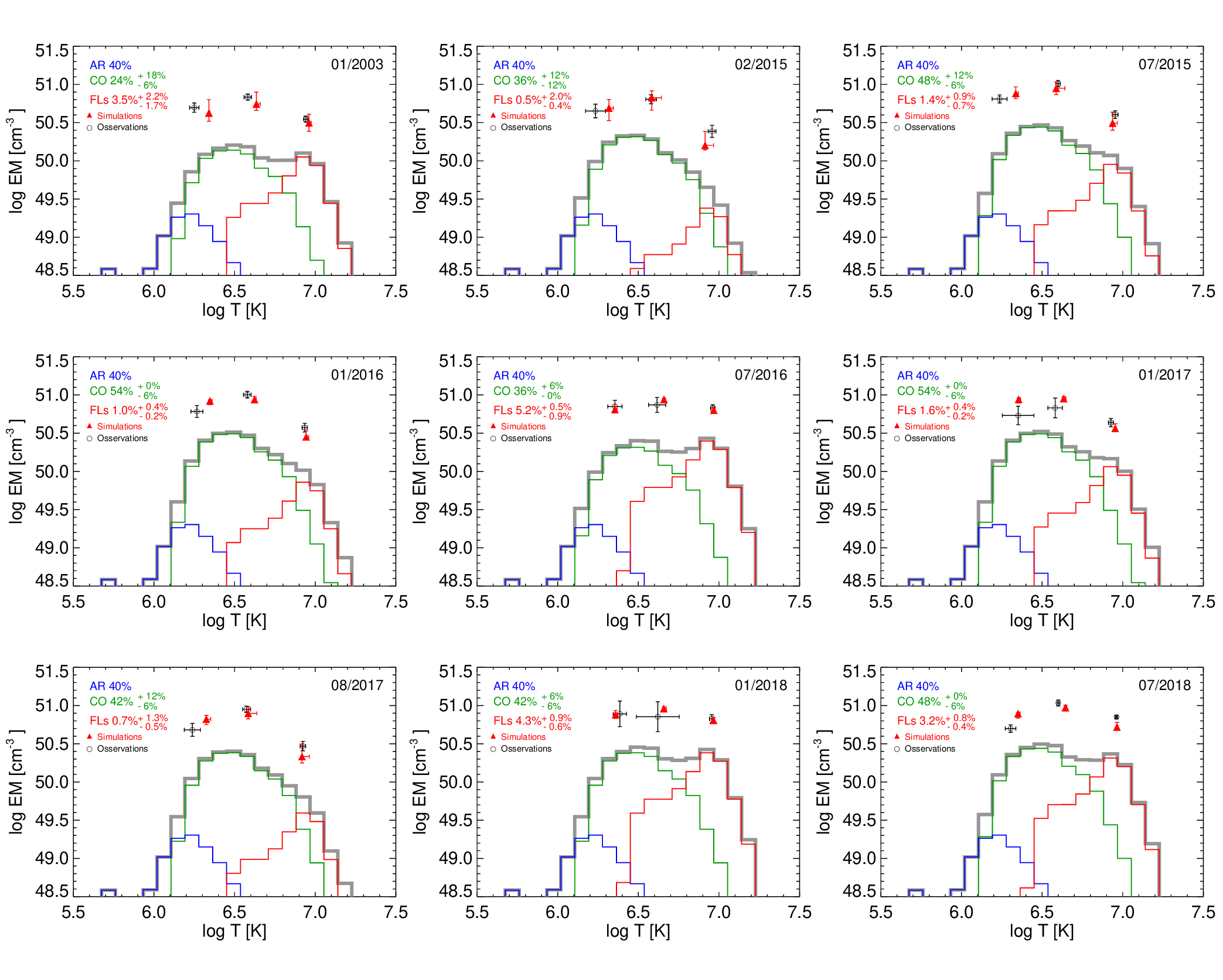}
\caption{Selected best-fitting EMD for class M flares at the soft and hard energies during the decay phase. The coding of the plots is the same as in \ref{fig:1stEMD}.}
\label{fig:Msoft_hard_decay}
\end{figure*}

\begin{figure*}
\centering
\includegraphics[width=\hsize]{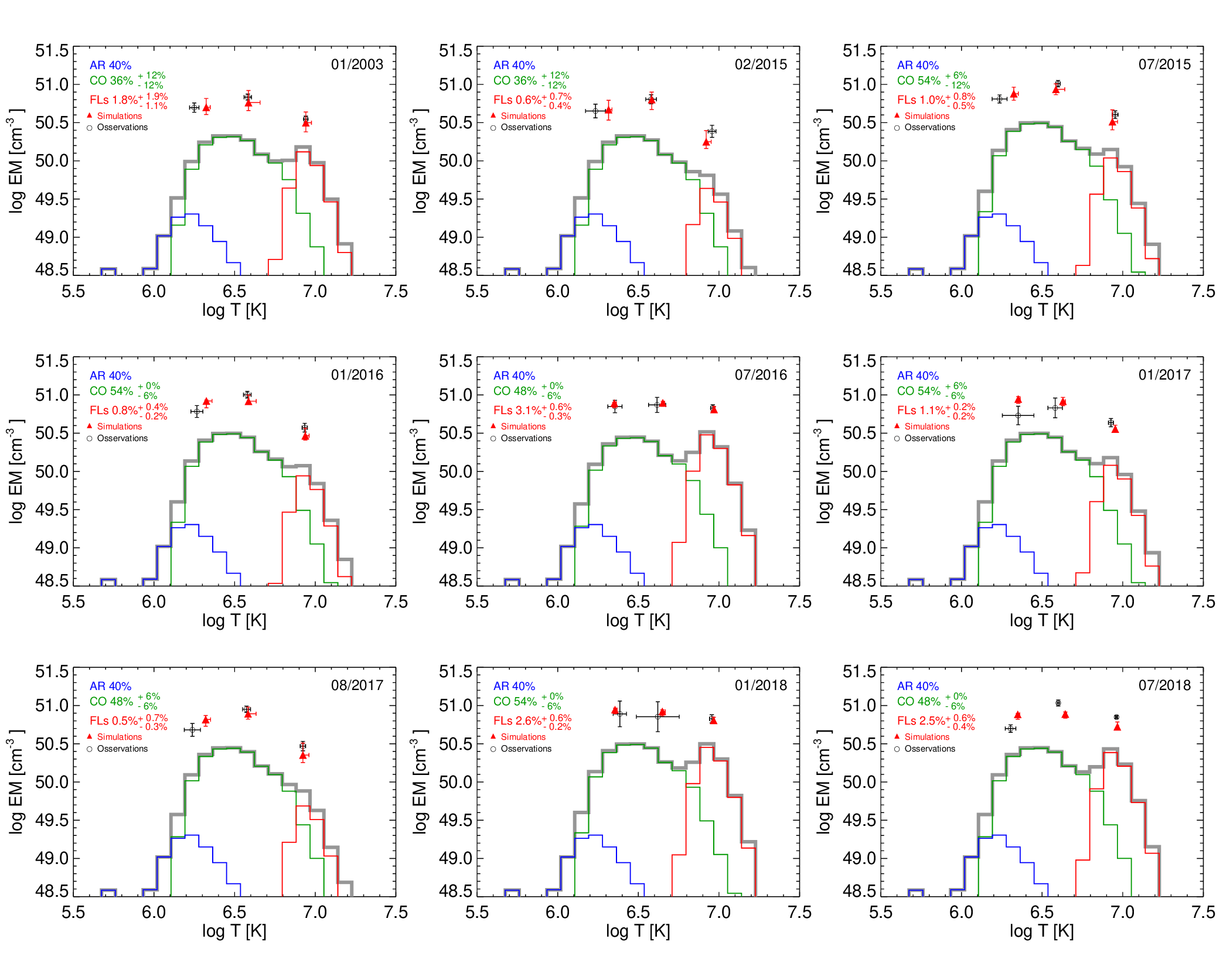}
\caption{Selected best-fitting EMD for class M flares at the hard energies during the decay phase. The coding of the plots is the same as in \ref{fig:1stEMD}.}
\label{fig:Msoft_decay}
\end{figure*}

\end{appendix}

\end{document}